\documentclass[aps,prl,superscriptaddress,groupedaddress,floatfix,longbibliography,twocolumn]{revtex4-2}
\setcitestyle{super}

\usepackage[en-US,useregional,showdow]{datetime2}
\DTMlangsetup[en-US]{ord=raise}

\usepackage{amsthm}
\theoremstyle{definition}

\theoremstyle{plain}

\usepackage{graphicx}
\usepackage{longtable}
\usepackage{wrapfig}
\usepackage{rotating}
\usepackage[normalem]{ulem}
\usepackage{amsmath}
\usepackage{amssymb}
\usepackage{capt-of}
\usepackage{hyperref}
\usepackage{amsthm}
\usepackage{xcolor}
\usepackage{physics}
\usepackage{slashed}
\usepackage{siunitx}
\usepackage{mathtools}
\usepackage[capitalize]{cleveref}
\Crefname{equation}{Eq.}{Eqs.}

\usepackage[inline]{enumitem}
\newlist{captionlist}{enumerate*}{2}
\setlist[captionlist,1]{label=\textbf{(\alph*)}}
\setlist[captionlist,2]{label=\textbf{(\alph{captionlisti}.\roman*)}}

\let\oldsout\sout
\renewcommand{\sout}[1]{{\color{red}{\oldsout{#1}}}}

\newcounter{para}
\newcommand{\para}{\par\refstepcounter{para}\textbf{{\color{cyan}[\thepara]}}\space}
\let\para\relax

\hypersetup{
 pdfauthor={Yuri D. Lensky},
 pdftitle={Mobile non-Abelian anyons in a qubit stabilizer code},
 pdfkeywords={},
 pdfsubject={},
 pdfcreator={}, 
 pdflang={English}}

\DeclareMathOperator{\pathverts}{vertices}
\DeclareMathOperator{\pathlen}{length}

\makeatletter
\newcommand{\restore@Environment}[1]{%
  \AtBeginDocument{%
    \csletcs{#1*}{#1}%
    \csletcs{end#1*}{end#1}%
  }%
}
\forcsvlist\restore@Environment{alignat,equation,gather,multline,flalign,align}
\makeatother

\let\autocite\cite

\newcommand{\xCornell}{Department of Physics, Cornell University, Ithaca, NY, USA}
\newcommand{\xEwha}{Department of Physics, Ewha Womans University, Seoul, South Korea}
\newcommand{\xHarvard}{Department of Physics, Harvard University, Cambridge, MA, USA}
\newcommand{\xRadcliffe}{Radcliffe Institute for Advanced Studies, Cambridge, MA, USA}
\newcommand{\xGoogle}{Google Research, Mountain View, CA, USA}

\begin{document}
\author{Yuri D. Lensky}
\affiliation{\xCornell}
\author{Kostyantyn Kechedzhi}
\affiliation{\xGoogle}
\author{Igor Aleiner}
\affiliation{\xGoogle}
\author{Eun-Ah Kim}
\affiliation{\xCornell}
\affiliation{\xRadcliffe}
\affiliation{\xHarvard}
\affiliation{\xEwha}
\date{\today}
\title{Graph gauge theory of mobile non-Abelian anyons in a qubit stabilizer code}

\begin{abstract}
  Stabilizer codes allow for non-local encoding and processing of quantum
  information. Deformations of stabilizer surface codes introduce new and
  non-trivial geometry, in particular leading to emergence of long sought
  after objects known as projective Ising non-Abelian anyons. Braiding of
  such anyons is a key ingredient of topological quantum computation.  We
  suggest a simple and systematic approach to construct effective unitary
  protocols for braiding, manipulation and readout of non-Abelian anyons and
  preparation of their entangled states.  We generalize the surface code to a
  more generic graph with vertices of degree 2, 3 and 4.  Our approach is
  based on the mapping of the stabilizer code defined on such a graph onto a
  model of Majorana fermions charged with respect to two emergent gauge
  fields.  One gauge field is akin to the physical magnetic field. The other
  one is responsible for emergence of the non-Abelian anyonic statistics and
  has a purely geometric origin. This field arises from assigning certain
  rules of orientation on the graph known as the Kasteleyn orientation in the
  statistical theory of dimer coverings. Each 3-degree vertex on the graph
  carries the flux of this ``Kasteleyn'' field and hosts a non-Abelian anyon.
  In our approach all the experimentally relevant operators are unambiguously
  fixed by locality, unitarity and gauge invariance.  We illustrate the power
  of our method by making specific prescriptions for experiments verifying the
  non-Abelian statistics.
\end{abstract}

\maketitle

\para Topological quantum
computation\cite{kitaevFaulttolerantQuantumComputation2003,nayakNonAbelianAnyonsTopological2008,sternNonAbelianStatesMatter2010}
 can be realized by a macroscopic quantum system with a few controllable collective degrees of freedom, called non-Abelian anyons.
Multiple non-Abelian anyons define a Hilbert space, whose dimension is set by the number and type of non-Abelian anyons.
States in this Hilbert space encode information non-locally. Hence they can serve as a quantum memory
protected from local perturbations.  Quantum gates that process this quantum
information are to be implemented through exchanges of pairs of anyons
that braid their space-time trajectories {[see Fig. 1(a)]}.  A double
braiding of identical non-Abelian anyons, an exchange of the positions of a
pair of anyons twice that returns them to a locally indistinguishable state,
may nonetheless change physical observables of the system.  Since the
braiding outcome of non-Abelian anyons are insensitive to details of the
anyon trajectories the implementation of quantum gates by braiding
non-Abelian anyons are topologically protected.

\para A simple construction of non-Abelian anyons is based on Majorana
fermions \(\alpha_j\), satisfying \(\{\alpha_j, \alpha_k\} = 2 \delta_{jk}\). Two Majorana operators
define a parity for a complex fermion with number \(n\),
\(i \alpha_2 \alpha_1 = (-1)^{n}\). Separating them in space is sufficient to realize
quantum memory. We now describe how Ising
non-Abelian\cite{mooreNonabelionsFractionalQuantum1991a,nayak2nquasiholeStatesRealize1996,kitaevAnyonsExactlySolved2006}
braiding arises for Majorana fermions bound to \(\pi\) flux, following an
argument of \textcite{ivanovNonabelianStatisticsHalfquantum2001} for the case
of \(p + ip\) superconductors. Consider a system of four Majoranas,
$\alpha_i, i=1,..,4$, in {Fig.~1(b-c)}. Bringing Majoranas together allows local
measurement of the fermion parity $i\alpha_2\alpha_1$. Double braiding of
$\alpha_3$ and $\alpha_2$ {[see Fig.~1(c)]} is equivalent to moving
$\alpha_2$ around $\alpha_3$ {[see Fig.~1(b)]}. Since Majorana $\alpha_3$ carries
$\pi$ flux and the Majorana $\alpha_2$ carries charge, the latter picks up a phase
$\alpha_2 \rightarrow -\alpha_2$ similarly to Aharonov-Bohm effect. Therefore the fermion parity
$i\alpha_2\alpha_1$ changes
sign.
Hence this double braiding results in a rotation in the Hilbert space of
anyons. In other words, if $i\alpha_2\alpha_1$ is identified with a Pauli $Z$
operator,
the braiding realizes an $X$ logical gate.  However, despite decades of
research\cite{readPairedStatesFermions2000a,
  ivanovNonabelianStatisticsHalfquantum2001,
  mooreNonabelionsFractionalQuantum1991a,
  bondersonDetectingNonAbelianStatistics2006,
  clarkeMajoranaFermionExchange2011a,
  aliceaNonAbelianStatisticsTopological2011b,
  lindnerFractionalizingMajoranaFermions2012a,
  sternTopologicalQuantumComputation2013a, clarkeExoticNonAbelianAnyons2013}
non-Abelian anyons were never unambiguously
  observed in experiment.

\begin{figure*}[htbp]
  \centering
  \includegraphics[width=\textwidth]{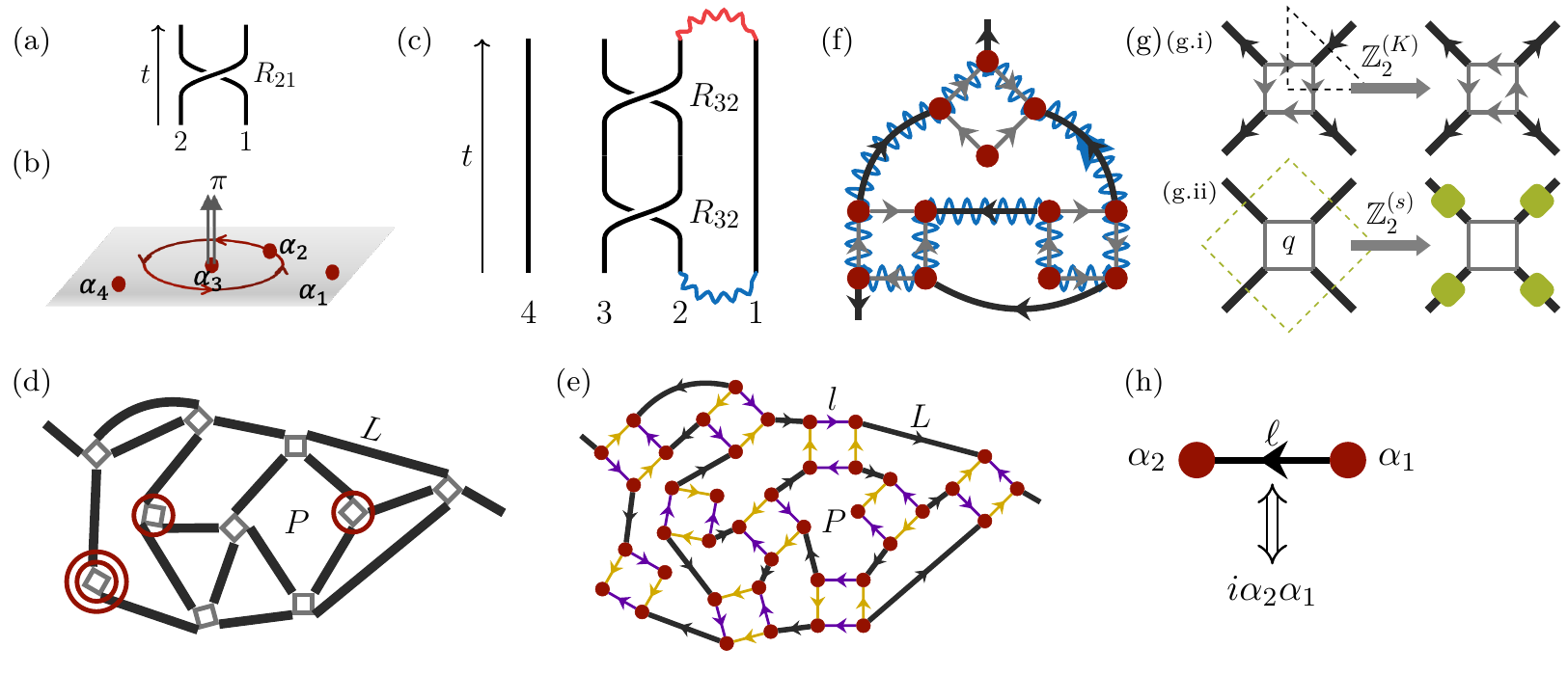}
  \caption[]{%
    \begin{captionlist}
      \item A schematic of the counter-clockwise swap $R_{12}$ of two anyons 1 and 2.
      \item  Aharonov-Bohm effect of flux-bound Majorana fermions.
      \item Double braid of anyons 2 and 3. The wavy blue/red line show fermion parity operator defined before/after the double braiding operation.
      \item Plaquette surface code graph with qubits drawn as gray diamonds. D3Vs are marked with red circles and a D2V is marked with a double-circle.
      \item The decorated version of the PSC graph in Fig. 1 (d) with Kasteleyn orientation. Red dots correspond to Majorana fermions. Black links connect different qubits, and yellow and purple links are intra qubit.
      \item An example counter-clockwise canonical loop enclosing a single \(\sigma\).
       \item The two emergent \(\mathbb{Z}_2\) gauge symmetries in our model. Both local
      symmetry actions correspond to (contractible) loops in the dual
      graph shown in dashed lines.
      \begin{captionlist}
        \item the \(\mathbb{Z}_2^{(K)}\) symmetry transformation from one Kasteleyn orientation to another flips all arrows touching a vertex.
        \item the \(\mathbb{Z}_2^{(s)}\) symmetry transformation generated by \(\Gamma_q\) flips all \(L\)-type links touching the diamond $q$.
      \end{captionlist}
        \item A diagrammatic rule for assigning directed edges to Majorana bilinears on the $\ell$-edges. 
    \end{captionlist}%
  }
\end{figure*}

\para Recent development of gate based quantum
processors\cite{satzingerRealizingTopologicallyOrdered2021} provides a new
avenue for direct preparation of a many-body quantum state without involving
the Hamiltonian and the difficulty in reaching its ground state. We introduce
the plaquette surface code (PSC) as a stabilizer
code\cite{gottesmanStabilizerCodesQuantum1997} defined on a specific type of
{\it qubit graph} [see Fig. 1(d)]. As in any stabilizer code, the multi-qubit
state $\ket{\psi}$ can be prepared to satisfy commuting constraints,
\begin{equation}
  B(P)\ket{\psi}=\ket{\psi},\label{eq:B(P)}
\end{equation}
where $B(P)$ are operators called \textit{stabilizers} for each plaquette $P$
of the qubit graph [see Fig. 1(d)]. The states $\ket{\psi}$ satisfying
Eq.~\eqref{eq:B(P)} form the {\it code subspace}. A state $\ket{\phi}$ with
\(B(P') \ket{\phi} = - \ket{\phi}\) for a plaquette \(P'\) has a {\it ``stabilizer
  flux''} at \(P'\). In the rest of the paper, we focus on states with few to
no stabilizer fluxes.  In the PSC, the qubits form vertices of a surface
graph, which only contains degree 4 (D4Vs), degree 3 (D3Vs), and degree 2
(D2Vs) vertices {[See Fig.~1(d)]}. We will show D3V's host Ising anyons.
  
\para The standard surface codes (on manifolds with and without boundary)
\cite{kitaevFaulttolerantQuantumComputation2003,wenQuantumOrdersExact2003a,bravyiQuantumCodesLattice1998,freedmanProjectivePlanePlanar1998}
are a special case of the PSC. \textcite{kitaevAnyonsExactlySolved2006}
pointed out the topological degrees of freedom at dislocations of the square
lattice, and \textcite{bombinTopologicalOrderTwist2010} and
\textcite{kitaevModelsGappedBoundaries2012a} pointed out that such
dislocations act as non-Abelian Ising anyons when they are introduced to the
toric code ground state\cite{kitaevFaulttolerantQuantumComputation2003}. This
observation motivated efforts to exploit the projective non-Abelian nature
\cite{barkeshliTheoryDefectsAbelian2013,youProjectiveNonAbelianStatistics2012a,benhemouNonAbelianStatisticsMixedboundary2021}
of the so-called ``twist defects'' which were found to carry Majoranas
\cite{zhengDemonstratingNonAbelianStatistics2015,brownPokingHolesCutting2017a}. 
However, the microscopic mechanism of flux attachment was not identified and
a protocol for moving the defects unitarily is absent.  Moreover,
manipulation of anyons can be realized by code deformation,
i.e. reconfiguration of the stabilizers and the movement of the edges of the
graph. In absence of the microscopic gauge theory, the design of optimal anyon
manipulation protocols is challenging. The operational use of the graph in
our approach is to define directed paths. Those directed paths enable us to
simply and systematically find all essential operators: the stabilizers,
unitary operators for dynamics, and Hermitian operators for the logical qubit
state measurements.

\para In this paper, we explicitly identify a gauge field responsible for the
flux attachment on a graph, and demonstrate its purely geometric origin.  By
formulating a new graph gauge theory, we construct optimal unitary protocols
for projective Ising anyon state preparation and braiding, and predict
specific experimental outcomes.  Note that the surface codes were recently
implemented on gate based NISQ superconducting
processors\cite{satzingerRealizingTopologicallyOrdered2021,acharyaSuppressingQuantumErrors2022}. Our unitary protocols are advantageous for such platforms
since for them unitary operations are typically faster than measurement based
protocols by an order of magnitude.

\para As usual the gauge field is associated with a global conserved
quantity. On any graph \(G\) where all vertices are of degree 2, 3, and 4,
the number \(N_{\sigma} = N_{\text{D3V}} + 2 N_{\text{D2V}} = 0 \mod 2\), where
\(N_{\text{D}j\text{V}}\) is the number of degree \(j\) vertices, is
conserved \((\mathrm{mod} \; 2)\)\footnote{as a consequence of the
  ``handshaking lemma'' that every graph has an even number of odd degree
  vertices\cite{eulerSolutioProblematisAd1741}}. In fact, the value of
\(N_{\sigma}\) also has an important physical consequence and associated
conservation law: if there are \(N_S\) stabilizer plaquettes, Euler's formula
for the Euler characteristic \(\chi(M)\) yields
\begin{equation}
  \label{eq:qubit-stabilizer-count}
  N_Q - N_S = \frac{N_{\sigma}}{2} - \chi(M),
\end{equation}
where we take our surface graph on some manifold with boundary \(M\). From
this formula we find that the dimension of the code subspace in the most
important case, \(M\) topologically a disk, is
\(\max \{2^{N_{\sigma}/2 - 1}, 1\}\)\footnote{On a general manifold there may be
  relations amongst the stabilizers that depend on topology and boundary
  conditions which can increase the effective code subspace.}. This is the
first hint that each \(\sigma\) corresponds to non-local degrees of freedom, as
each is roughly ``half'' a qubit\footnote{In other words, an anyon with the quantum dimension $\sqrt{2}$.}. Importantly, if the number of stabilizers
is fixed, \(N_{\sigma}\) is conserved.

\para To make the conservation of \(N_{\sigma}\) more manifest, we decorate each
qubit vertex with a diamond as shown in {Fig.~1(e)}. On the decorated graph
\(\tilde{G}\), \(N_{\sigma}\) is the number of vertices with two incident edges,
which we call \(\sigma\) or ``unpaired''. We construct a field which assigns flux
to these vertices in a particular way.

\para First, we need a local rule to lift directed paths \(\gamma\) through the
``physical'' qubit graph \(G\) [Fig.~1(d)] to directed paths \(\tilde{\gamma}\) through
\(\tilde{G}\) [Fig.~1(e)]: every diamond is traversed counter-clockwise [Fig.~1(f)]. Such paths
\(\tilde{\gamma}\) are called canonical. Our field is the assignment of arrows
to each link, which follows the local rule that an odd number are clockwise
about each face (such an orientation is called
\emph{Kasteleyn}\cite{kasteleynDimerStatisticsPhase1963}\footnote{Kasteleyn
  structures were introduced to study dimer
  models\cite{kasteleynDimerStatisticsPhase1963}, and were later related to
  2D spin
  structures\cite{cimasoniDimersSurfaceGraphs2007,cimasoniDimersSurfaceGraphs2008}}). We
find [See Fig.~1(f) and Appendix B]
\begin{equation}
  (-1)^{N_\sigma(\tilde{\gamma})}= - \prod_{\ell \in \text{Edges}(\tilde{\gamma})} (-1)^{n^{(K)}_{\ell}(\tilde{\gamma})},
  \label{eq:N_sigma}
\end{equation}
for any counter-clockwise canonical loop \(\tilde{\gamma}\), where
\(n^{(K)}_{\ell}(\tilde{\gamma})\) is 1 (0) if the arrow on the edge \(\ell\) is in the
opposite (same) direction as \(\tilde{\gamma}\), and
\(N_{\sigma}(\tilde{\gamma})\) is the number of \(\sigma\) enclosed by the loop
\(\tilde{\gamma}\). The Kasteleyn orientation is not unique: for example, flipping
all the arrows touching a vertex is a local \(\mathbb{Z}_2\) transformation (which we
call \(\mathbb{Z}_2^{(K)}\)) from one Kasteleyn orientation to another [See
Fig.~1(g.i)], while manifestly preserving \Cref{eq:N_sigma} [See also
Appendix B]. In this sense, we have attached flux of a \(\mathbb{Z}_2\) field to the
\(\sigma\).

We now place a Majorana at each vertex of the decorated graph, Fig.~1(e). An
orientation is natural in a theory of Majorana fermions on a graph: after
assigning a direction, links \(\ell\) with an arrow \(\alpha_j \to \alpha_k\) define a
Hermitian fermion parity {[See Fig.~1 (h)]}
\begin{equation}
  (-1)^{n_{\ell}} = i \alpha_k \alpha_j.\label{eq:n_ell}
\end{equation}
The link operator is clearly not invariant under choice of particular
Kasteleyn structure. Since the \(\mathbb{Z}_2^{(K)}\) transformation at a vertex
hosting Majorana \(\alpha_j\) flips all the link operators involving
\(\alpha_j\), we can think of the Majoranas as ``charged'' under the local
\(\mathbb{Z}_2^{(K)}\) symmetry. If physical meaning could be given to canonical
paths, the Majoranas at \(\sigma\) vertices would be bound to \(\pi\) flux. We
describe a qubit model, the PSC, where there is both an emergent Kasteleyn
structure as well as a second \(\mathbb{Z}_2\) field associated to a gauge
transformation we call \(\mathbb{Z}_2^{(s)}\) [See Fig.~1(g.ii)]. Keeping the second field
flat ensures Wilson lines in the gauge theory maintain a canonical form under
local unitary evolution. Moreover, since no physical observable depends on
the particular Kasteleyn orientation chosen, in other words
\(\mathbb{Z}_2^{(K)}\) is gauged,
the Majoranas at \(\sigma\) vertices are bound to \(\pi\) flux of a gauge field.

\para We start by using the Kasteleyn orientation on the decorated graph to
explicitly determine two standard elements defining a gauge theory: the
physical subspace of the Majorana Hilbert space (giving rise to
\(\mathbb{Z}_2^{(s)}\)), and the mapping from physical qubits into the
subspace. Recall we placed a Majorana at each vertex of the decorated graph, so
that each qubit \(q\) of the PSC corresponds to a diamond with 4
Majoranas. Note that at each diamond, opposite links \(l_a, l_b\) do not
touch, so the operators [See Fig.~2(a)] \(\tau_1 = (-1)^{n_{l_a}}, \tau_1' = (-1)^{n_{l_b}}\)
satisfy \(\tau_1^2 = \tau_1'^2 = 1, [\tau_1, \tau_1'] = 0\), and neither can be
proportional to \(1\) since they anti-commute with the other pair of link
operators. In a \emph{qubit} Hilbert space, these conditions imply that
\(\tau_1 = \pm \tau_1'\), and the choice \(\tau_1 = \tau_1'\) in the qubit space gives rise
to the physical subspace condition
[see Fig.~2(b)]
\begin{equation}
  \label{eq:gauge-constraint}
  \Gamma_q \ket{\psi} = \ket{\psi}, \; \Gamma_q = (-1)^{n_{l_a}} (-1)^{n_{l_b}}.
\end{equation}
The Kasteleyn condition ensures that as an operator \(\Gamma_q\) is independent of
the chosen pair of edges, so if we construct \(\tau_2, \tau_2'\) in an analogous
way for the other pair we also find \(\tau_2 = \tau_2'\) in the physical
subspace. \(\Gamma_q\) generates a local gauge transformation
\(\mathbb{Z}_2^{(s)}\) under which each Majorana fermion carries a charge,
i.e. \(\alpha_{qj}\) changes sign upon conjugation with \(\Gamma_q\). The second
ingredient of the gauge structure, a mapping from qubits to the Majoranas, is
fixed\footnote{\label{fn:globalphase}More precisely, up to a global phase, which for us is irrelevant.} by choosing a qubit
operator to correspond to each pair of opposing edges, e.g.  Pauli operators
\(\tau_1 = Z\) and \(\tau_2 = X\). Note that, by construction, the spin operators
defined by the \(l\)-type links, \(\tau_1, \tau_2\), are invariant under
\(\mathbb{Z}_2^{(s)}\) and \(\mathbb{Z}_2^{(K)}\).

\para \emph{Stabilizers and \(\mathbb{Z}_2^{(s)}\) --} The final ingredient to define
our gauge theory is a local flatness condition for the \(\mathbb{Z}_2^{(s)}\) gauge
field formed by the inter-diamond \(L\)-type link operators\footnote{Note that we have
  given the analogous condition, an odd number of clockwise arrows in each
  plaquette, for the Kasteleyn orientation.}. In contrast to the intra-diamond \(l\)-type
link operators, which are \(\mathbb{Z}_2^{(s)}\)-invariant, \(L\)-type link operators
all commute (since these links never touch) but are odd under both
\(\mathbb{Z}^{(s)}_2\) and \(\mathbb{Z}_2^{(K)}\) [See Fig.~1(g)]. Specifically, the
\(\mathbb{Z}_2^{(s)}\) transformation flips all \((-1)^{n_L}\) touching a diamond. The
simplest \(\mathbb{Z}_2^{(s)}\)-invariant combination is a loop of \(L\)-type edges
around a stabilizer plaquette \(P\),
\begin{equation}
  \label{eq:gauge-bp-definition}
  B(P) = \prod_{L \in P} (-1)^{n_L}.
\end{equation}
Moreover, by writing \(B(P)\) in terms of the gauge-invariant \(l\)-type link
operators (this is a special case of \Cref{eq:aW-pauli}), we find that it is
\(\mathbb{Z}_2^{(K)}\) invariant as well. This gives both the definition of and
physical meaning to the stabilizers defining the PSC code subspace alluded to
in \Cref{eq:B(P)}.

\para \emph{Emergence of \(\mathbb{Z}_2^{(K)}\) --} Since the Kasteleyn orientation is
not a conventional \(\mathbb{Z}_2\) gauge field, let us briefly describe an
alternative construction of the same theory where the gauge structure is
emergent. A consistent mapping from the single qubit Hilbert space into a
fixed parity sector of 4 Majoranas is fully
specified~\cite{Note5}
by associating a diamond with Kasteleyn orientation to the qubit, and pairs
of opposite edges on the diamond to two generators of the Pauli algebra, as
above. Extending this construction to a multi-qubit system, by additionally
assigning arrows to \(L\)-type links the corresponding (gauge-non-invariant
and hence unphysical) operators combine to measure a (physical)
\(\mathbb{Z}_2^{(s)}\) gauge flux \Cref{eq:gauge-bp-definition}. If the arrows are
assigned so that \emph{every} plaquette has a Kasteleyn orientation, \(B(P)\)
is simply a product of Pauli operators at each diamond determined by the
local embedding at each qubit of the plaquette, regardless of the size or
shape of \(P\).\footnote{In general, the same would be true if we fixed the
  parity of clockwise edges across all stabilizer plaquettes.} We note that a
static graph with a preferred mapping between qubits and Majoranas dictated
by a Hamiltonian, as in the model studied by
\textcite{kitaevAnyonsExactlySolved2006}, may fix part of the Kasteleyn
structure. However, as D3Vs and D2Vs move, the PSC
evolves. In this case, the emergent $\mathbb{Z}_2^{(K)}$ plays a critical role in
tracking the PSC evolution.

\para Having defined the complete gauge theory, we consider two families of
multi-qubit operators that act on the PSC state, distinguished by the
condition that they generate stabilizer flux only at controlled
locations\cite{fradkinFieldTheoriesCondensed2013}\footnote{More precisely, we
  will find violations of different type are created in pairs at the ends of
  string operators.}. Acting with a Majorana on a given vertex flips the edge
operators \((-1)^{n_{\ell}}\) for every edge \(\ell\) touching the vertex, creating
a pair of stabilizer fluxes if the vertex is not unpaired [See Fig.~2(c.i)].
The local condition not to create stabilizer flux is to flip an \emph{even}
number of \(L\)-type edge operators around each plaquette: either acting with
Majoranas on both ends of an $L$-type edge, i.e., $(-1)^{n_L}$ [See
Fig.~2(c.ii)], or to flip 2 or more \(L\)-type links about each stabilizer
plaquette [See Fig.~2(c.iii)]. Combined with local gauge invariance, the
first method builds Wilson lines\cite{wilsonConfinementQuarks1974}, while the
second builds 't Hooft lines\cite{thooftPhaseTransitionPermanent1978}.

\begin{figure*}[htbp]
  \centering
  \includegraphics{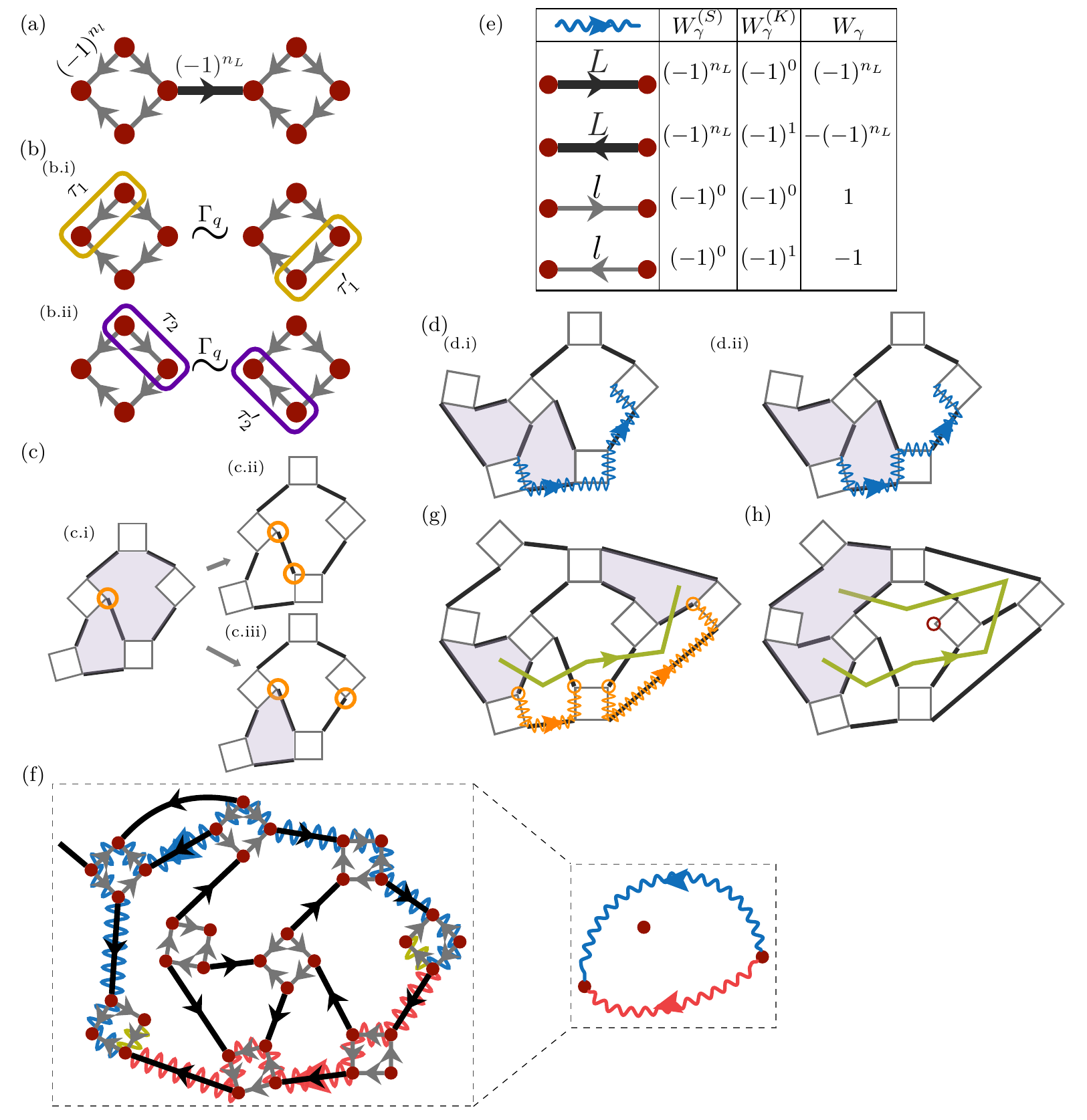}
  \caption[]{%
    \begin{captionlist}
      \item The parity operators on the $l$-edges and $L$-links.
      \item Pauli operator $\tau_1, \tau_2$ assignment to the diamond edges $ \{\tau_1, \tau_2\}=0$. For instance, $\tau_1=Z$ and $\tau_2=X$. 
      \item Two examples of how to locally avoid creating stabilizer flux,
      ignoring gauge invariance.
      \begin{captionlist}
        \item Acting with the  Majorana circled in orange flips the \(L\)-link it
      touches, and creates \(\varepsilon\), i.e., a pair of stabilizer fluxes sharing an edge, shown in lavender.
        \item The basis for the Wilson line. Flip the same $L$-link again by acting with a second Majorana touching the link.
        \item The basis for the 't Hooft line. Flip an even number of $L$-links. 
      \end{captionlist}
      \item Augmented Wilson lines:
      \begin{captionlist}
        \item An example of a canonical augmented Wilson line action on a state with no
        stabilizer flux. The link at the start of the line is flipped, creating an
        \(\varepsilon\), while there is no bond to flip at the unpaired vertex. If there were
        an \(\varepsilon\) particle at the start of the line, this augmented Wilson line
        would ``sink'' it into the unpaired vertex, removing any stabilizer flux.
        \item A non-canonical Wilson line.
      \end{captionlist}
      \item The diagrammatic rules for constructing a Wilson line operator from a
      directed path on the graph. The blue wavy arrow indicates the orientation of the path $\gamma$. The black and gray arrows indicate Kasteleyn orientation on the $L$-links and on the $l$-edges respectively. 
      \item Two canonical augmented Wilson lines between the same pair of unpaired vertices. We show the common part of the two lines, which in this case lie on the first and last links, in dark yellow.
      \item An open 't Hooft line is shown in moss green, with stabilizer fluxes at its ends. The shown 't Hooft line is equivalent to the gauge-invariant Wilson line segments shown in orange.
      \item An  open 't Hooft line wrapping an unpaired vertex, marked with a red circle, creating an \(\varepsilon\) from a state with no
      stabilizer flux.
    \end{captionlist}%
  }
\end{figure*}

\para \emph{Wilson lines --} Flipping each \(L\)-type link twice means we act
with \(L\)-link operators, which manifestly commute with \(B(P)\).  While
\((-1)^{n_{L}}\) is not gauge invariant under \(\mathbb{Z}_2^{(s)}\), if we chain
\(L\)-link operators (connected by diamonds), the bulk of the chain commutes
with \(\Gamma_q\).  To make the ends of the chain \(\mathbb{Z}_2^{(s)}\) invariant, we must
add an additional Majorana from the diamonds at the ends, arriving at the
definition of a \emph{valid} path for the augmented Wilson line in
Fig.~2(d). Formally, a valid path is one that starts and ends on
\(l\)-links. To give a definition of an operator that is both consistent with
the Majorana anti-commutation relations and invariant under
$\mathbb{Z}_2^{(K)}$, we take the path \(\gamma\) (from
\(\alpha_I \to \alpha_F\)) to be directed. Explicitly, the gauge-invariant ``augmented
Wilson line'' associated to the path \(\gamma\) is defined by [See Fig.~2(e)]
\begin{align}
  \mathcal{W}_{\gamma} &= i \alpha_F W_{\gamma} \alpha_I, \; W_{\gamma} = W^{(s)}_{\gamma} W^{(K)}_{\gamma} \label{eq:aW}\\
  W^{(s)}_{\gamma} &= \prod_{L\in\gamma}(-1)^{n_{L}}, \; W^{(K)}_{\gamma} = \prod_{\ell \in \gamma} (-1)^{n^{(K)}_{\ell}(\gamma)},
                \label{eq:WsK}
\end{align}
where we refer to \(W_{\gamma}\) as the Wilson line.  If the line is open, its
ends are either paired or unpaired vertices. If the vertex is paired, a pair
of stabilizer fluxes sharing an edge are created when acting on a state with
no stabilizer flux: we call such flux configurations an
\(\varepsilon\)-particle\cite{kitaevFaulttolerantQuantumComputation2003} {[See
  Fig.~2(d)]}. No stabilizer flux is created at an unpaired end.

\para \emph{Wilson loops and line deformations --} When \(\gamma\) is a (directed)
loop the Wilson loop \(W_{\gamma}\), which can be defined by the same
\Cref{eq:aW,eq:WsK}, is gauge invariant on its own. We define the augmented
Wilson loop (\Cref{eq:aW} is not defined for coinciding ends) as
\(\mathcal{W}_{\gamma} = - W_{\gamma}\) (to emphasize the type we sometimes write
\(\mathcal{W}^{(\text{loop})}_{\gamma}\)). A canonical counter-clockwise augmented Wilson
loop measures the parity of stabilizer and Kasteleyn flux:
\begin{equation}
  \label{eq:augmented-wilson-loop-flux}
  \mathcal{W}^{(\text{loop})}_{\gamma} = (-1)^{N_{\Phi}(\gamma)} (-1)^{N_{\sigma}(\gamma)},
\end{equation}
where \(N_{\Phi}(\gamma)\) is the operator measuring the stabilizer flux enclosed by
the loop. It is practically useful that the operator \(B(P)\) is just the
counter-clockwise augmented Wilson loop about only the stabilizer plaquette
\(P\): one perspective is that \(B(P)\) should only count the stabilizer flux
\(B(P) = (-1)^{N_{\Phi}(P)}\), so is not equivalent to a canonical loop in the
presence of anyons. The most important application is to the ratio of
canonical Wilson lines for two paths \(\gamma, \gamma'\) between same anyons 1 and 2
{[See Fig.~2(f)]}. The gauge-invariant operator
\(\mathcal{W}_{\gamma} \mathcal{W}_{\gamma'}^{-1} = W_{\gamma} W_{\gamma'}^{-1}\) can be decomposed to a product of
canonical augmented Wilson loops, such that for canonical paths
\begin{equation}
  \label{eq:wilson-line-deformation}
  \mathcal{W}_{\gamma} \mathcal{W}_{\gamma'}^{-1} = W_{\gamma} W_{\gamma'}^{-1} = (-1)^{N_{\Phi}(\gamma,\gamma')} (-1)^{N_{\sigma}(\gamma,\gamma')}.
\end{equation}
where \(N_{\Phi}(\gamma,\gamma')\) is the stabilizer flux enclosed between the paths,
whereas \(N_{\sigma}(\gamma,\gamma')\) is the number of enclosed anyons (see Appendix B.1
for the precise definition, which is only needed when one of the paths
\(\gamma,\gamma'\) goes directly through a diamond containing an anyon away from the
endpoints of \(\gamma,\gamma'\), and therefore does not play an important role in
braiding).

\para \emph{'t Hooft lines and loops --} 
We represent an 't Hooft line\cite{thooftPhaseTransitionPermanent1978} as a
directed path of \emph{even} length through the dual graph, whose links
represent the flipped \(L\)-type bonds [See Fig.~2(g)]. The definition
ensures that we can always find a local, gauge-invariant operator
corresponding to the 't Hooft line.  Specifically, an 't Hooft line can be
written as a product of augmented Wilson lines by taking the Majoranas to the
right of the path which touch the links crossed, and making this product
gauge invariant in the most local way {[See Fig.~2(g)]}.  If the path is
open, stabilizer fluxes are created at its ends. Flips corresponding to odd
length paths through the dual lattice are always products of 't Hooft lines
and an augmented Wilson line with one unpaired end.  Finally, we note that 't
Hooft loops create no flux; for simplicity, we always take such loops
counter-clockwise.

\para \emph{'t Hooft and canonical Wilson lines --} Two relationships between
the two families of operators are of particular importance.  First, note that
a Wilson and 't Hooft line anti-commute at each point of crossing, because 't
Hooft lines flip \(L\)-type links.  Since \(\varepsilon\) live at the end of augmented
Wilson lines {[See Fig.~2(d)]}, roughly speaking 't Hooft loops detect
the parity of enclosed \(\varepsilon\). More importantly, certain 't Hooft lines going
around a single anyon counter-clockwise are equivalent, in a no-flux state,
to \emph{canonical} Wilson lines [See Fig.~3(a)]. Specifically, an 't Hooft line going around
a single anyon counter-clockwise cannot be closed to a non-intersecting
loop. The ends \emph{can} be brought to adjacent
plaquettes, where an \(\varepsilon\) will be created {[See Fig.~2(h)]}. This acts in the same way as an
augmented Wilson line starting at the anyon and ending at the
\(\varepsilon\). As we demonstrate in  {Fig.~3(a)}, the augmented Wilson line that
has the same action \emph{including the global phase} can always be taken to
follow a canonical path.

\begin{figure*}[htbp]
  \centering
  \includegraphics[width=\textwidth]{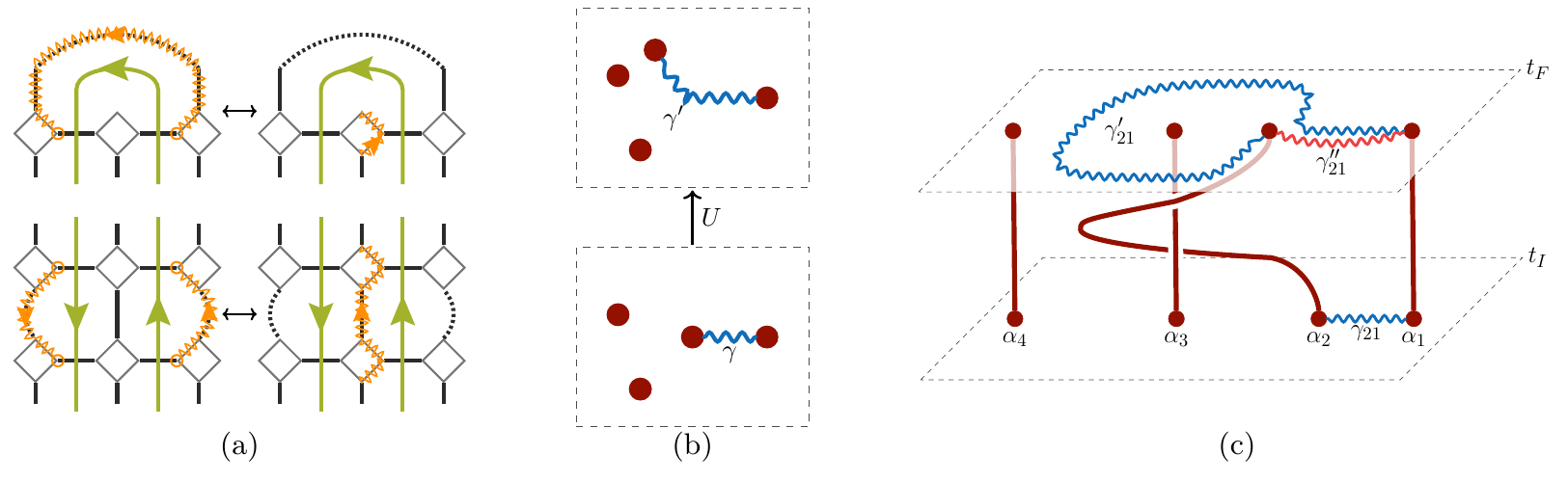}
  \caption[]{%
    \begin{captionlist}
    \item Examples of an explicit deformation of 't Hooft lines (moss green) to augmented Wilson line (orange) (left)  and canonical Wilson lines (right), in a state with no stabilizer flux.
        \item Motion of the anyon $2$ generated by a local unitary \(U\), acting
      only on a region containing anyon $2$.
      \item A manifestly gauge invariant illustration of Ising anyon braiding in the long-distance limit, tracking the extending augmented Wilson line and the world line of the anyons.
      The augmented Wilson lines associated with 
      \(\gamma_{21}\) and \(\gamma_{21}''\) are measured before and after the double braid respectively. The comparison between the two measurements should only depend on  
      the topological form of the motion
      of \(\alpha_2\).
    \end{captionlist}%
  }
\end{figure*}

\para \emph{Extension of canonical lines --} General
principles of gauge theory\cite{wilsonConfinementQuarks1974} dictate that the
Wilson line \(W_{\gamma}\) between \(\alpha_1\) and \(\alpha_2\), associated to the
augmented Wilson line \(\mathcal{W}_{\gamma}\), should ``extend'' when
$\alpha_2$ moves by a local unitary $U$ starting from a state \(\ket{\psi}\). Specifically, locality, unitarity, and gauge invariance
determine the key aspects of the path \(\gamma'\) one should take \emph{after}
moving \(\alpha_2\) so that \(W_{\gamma'}\) ``acts the same way'' as
\(W_{\gamma}\). Explicitly, as shown in {Fig.~3(b)} the path \(\gamma'\) is just an
extension of \(\gamma\) into the region where \(U\) acts and ending at the new
location of \(\alpha_2\), so that
\(\mathcal{W}_{\gamma'} U \ket{\psi} = U \mathcal{W}_{\gamma} \ket{\psi}\). This way, the Wilson line keeps track
of the path and history of the anyons.

\para Therefore, the last key ingredient of our theory of non-Abelian anyons
is the requirement that Wilson lines are extended by motion. This is the
physical condition which distinguishes canonical Wilson lines: \emph{any}
local unitary \(U\) acting in a region \(A\) with a single anyon, without
stabilizer flux and preserving the anyon and stabilizer flux number, extends
any canonical Wilson line between anyons to a \emph{canonical} Wilson
line. In most cases, this can be seen by taking a canonical Wilson line
ending at the anyon and extending beyond \(A\), and (partially) expanding it
to an 't Hooft line that lies strictly outside \(A\); the first steps of this
expansion are shown on the right column of {Fig.~3(a)}. Acting with the
unitary cannot change the action of the 't Hooft line on the
state. Therefore, we can deform the 't Hooft line to a necessarily canonical
Wilson line ending at the new position of the anyon. Without this fundamental
property, the behaviour of Wilson lines would depend on non-topological
details of the dynamics. Instead, referring to
\Cref{eq:wilson-line-deformation}, we find that unpaired Majoranas carry both
a \(\pi\) flux and charge of the Kasteleyn field.

As discussed in the introduction, we can now conclude that the unpaired
Majoranas, or D3Vs, in our model are projective Ising anyons. To illustrate
this point directly, we formulate a simple braid to unambiguously demonstrate
non-Abelian statistics {(see Fig.~1(c))}. Initialize the system at time $t_I$
with four anyons arranged on a line, and suppose measurement of
\(\mathcal{W}_{\gamma_{21}}\) yields the value \(+1\) {[See Fig.~3(c)]}.  As we move
$\alpha_2$ \emph{around} $\alpha_3$, the path \(\gamma_{21}\) gets extended to a path
\(\gamma_{21}'\) around $\alpha_3$.  The measurement of
\(\mathcal{W}_{\gamma_{21}''}\) at time $t_F$ will give \(-1\), since it is different from
the measurement of \(\mathcal{W}_{\gamma_{21}}\) at time $t_I$ by
\(W_{\gamma_{21}'} W_{\gamma_{21}''}^{-1} = -1\) by \Cref{eq:wilson-line-deformation}
(\(N_{\Phi}(\gamma_{21}', \gamma_{21}'') = 0, N_{\sigma}(\gamma_{21}', \gamma_{21}'') = 1\)).  In other
words an observable changes sign after double braiding with probability 1,
which is sufficient to demonstrate non-Abelian statistics.  On the other
hand, if $\alpha_3$ was attached to a stabilizer flux the observable will not
change sign since now \(N_{\Phi}(\gamma_{21}', \gamma_{21}'') = 1\), so braiding about
such a composite could serve as a control experiment. We note that the
composites are on equal footing to what we consider the ``bare'' anyons, and
our notion of which anyon is a composite would switch if we had chosen to
prefer the opposite chirality, clockwise instead of counter-clockwise, in the
definition of the Kasteleyn structure and preferred loops. In particular,
amongst themselves the composites braid precisely as projective Ising anyons
as well.

\begin{figure*}[htbp]
  \centering
  \includegraphics{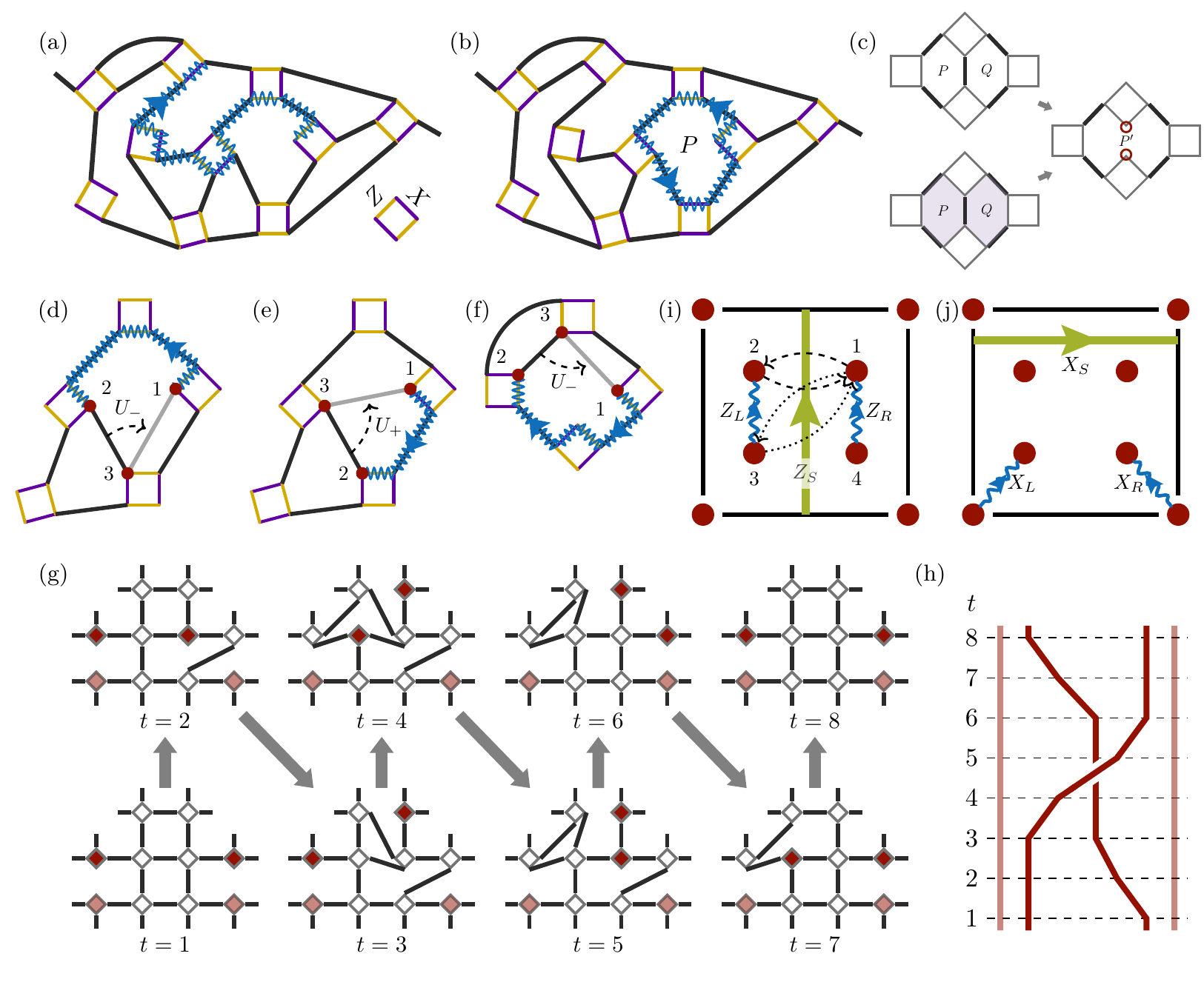}
  \caption[]{
    \begin{captionlist}
      \item A valid path defining an augmented Wilson line. The static assignement of the Pauli operators are indicated by yellow for $Z$ and purple for $X$. In this case, \(N_{ll}(\gamma)=3\) in \Cref{eq:aW-pauli}.
      \item A Wilson loop enclosing a stabilizer plaquette $P$. The Wilson loop operator graphically specifies the stabilizer \(B(P)\). 
      \item  Deleting an \(L\)-link in the graph creates two unpaired
      vertices marked with red circles. It also changes the stabilizers. If we take the new definition,
      \(B(P') \ket{\psi} = \ket{\psi}\) whether or not there was a fermion on
      \(P,Q\), without any unitary operation. This is a microscopic manifestation
      of the fusion rule \(\sigma \times \sigma = 1 + \varepsilon\).
      \item A basic edge move generated by a counter-clockwise augmented Wilson line shown in blue. Since the path is counter-clockwise, $U_-$ should be used to avoid creating \(\mathbb{Z}_2^{(s)}\) flux.  
      The vertex $3$, initially paired with $2$, pairs with $1$ after the edge move leaving $2$ unpaired.
      \item A basic edge move generated by a clockwise augmented Wilson line shown in blue. $U_+$ should be used to avoid creating \(\mathbb{Z}_2^{(s)}\) flux.
      \item Clockwise Wilson line with an anyon along the path. Since $N_{ll}(\gamma)=1$, one needs to use $U_-$ to avoid creating $\mathbb{Z}_2^{(s)}$ flux. 
      \item A protocol for implementing the braid group generator Fig.~1(a) and restoring the lattice.  We show a 10-qubit section for concreteness, but on a larger device the anyons could be further separated. 
      \item The time slices corresponding to the protocol in Fig.~4(g).
      \item The protocol for preparing a GHZ state of three logical qubits.  Four anyons at the corner and the two pairs created in the bulk together supply three logical qubits. The vertical 't Hooft line (moss green) and the two augmented Wilson lines (wavy blue) measures the logical qubit state. Dashed braiding or dotted braiding will both entangle the logical qubits to yield a GHZ state.
      \item The protocol for implementing the logical X operator.
    \end{captionlist}}
  \label{fig:fig-4}
\end{figure*}

\para Below we suggest specific protocols and predict outcomes for several experiments.

\para \emph{Spin operators for augmented Wilson lines and loops --} We will
use augmented Wilson lines and loops as the basis for all physical
operations, so it will only be necessary to give a qubit-space formula for
these operators. Fortunately, they can be constructed simply and
systematically from paths drawn on the decorated PSC graph (without assigning
any explicit Kasteleyn orientation). First, assign diamonds to each qubit,
and two Pauli generators, say \(X\) and \(Z\), to pairs of opposite edges on
each diamond {[Fig.~4(a)]}. In general, we call the Pauli associated to the
\(l\)-link \(\tau_{l}\), and we keep this assignment static. Now, \(L\)-type edges
are drawn between diamond vertices to construct a PSC graph. Given a valid
directed path \(\gamma\) in this graph, we simply read off the operator along the
path
\begin{equation}
  \mathcal{W}_{\gamma} = (-i)^{N_{ll}(\gamma)} \overleftarrow{\prod_{l \in \gamma}} \tau_{l}.
  \label{eq:aW-pauli}
\end{equation}
For multi-qubit loops \(\gamma\), we delete an $L$-link and apply
\Cref{eq:aW-pauli} to the resulting open path. Here \(N_{ll}(\gamma)\) is the
number of vertices in \(\gamma\) with adjacent \(l\)-edges {[See Fig.~4(a)]}. The
arrow over the product specifies that the product is to be taken in order
from right to left according to the path: $\tau_l$ for the earliest $l$ appears
at right end. 't Hooft lines are constructed using the correspondence to
products of augmented Wilson lines in Fig. 2 (g). We note that to make the
rules of the protocol simple, we will use both canonical and non-canonical
lines and loops.

\emph {Stabilizers and initial state -- } As an immediate application, we
recall that the stabilizers \(B(P)\) are simply the unique counter-clockwise
Wilson loops, generally not canonical, in the stabilizer plaquette \(P\)
{[See Fig.~4b]}. Hence, Eq.~\eqref{eq:aW-pauli} offers the necessary input for
a protocol to prepare a state in the code space of the PSC~\cite{satzingerRealizingTopologicallyOrdered2021}.

\para \emph{Creation, measurement, and fusion --} The creation of anyon pairs only requires the
removal of an \(L\)-type link {[See Fig.~4(c) top]}. When we modify the graph by deleting an
edge, we do not need to perform unitary action for the operators obtained on
the new graph to remain meaningful in the new code subspace. The link touches
at least one stabilizer plaquette \(P\). If the link is a boundary link, we
simply drop \(B(P)\) from the list of stabilizers. If the link touches
another plaquette, \(Q\), deleting the edge forms a larger plaquette \(P'\),
and we find \(B(P') = B(P) B(Q)\). Notice that if we remove a link shared by
the stabilizer fluxes of an \(\varepsilon\), we also end up in the no
stabilizer flux state of two additional anyons {[See Fig.~4(c) bottom]}. This embodies the Ising anyon
fusion rule\cite{kitaevAnyonsExactlySolved2006}
\begin{equation}
  \sigma \times \sigma = 1 + \varepsilon.
\end{equation}
Since arbitrary unitary motion preserves canonical Wilson
lines\footnote{Equivalently, since 't Hooft loops measure \(\varepsilon\) parity and
  deform to canonical augmented Wilson lines.}, if we wish to determine the
fusion state of separated anyons we should measure a canonical Wilson line
between them (according to the path along which they would be physically
fused). We could also measure the equivalent 't Hooft loop around the anyon
pair, but this is generally less efficient. We note that the other \(\sigma\)
fusion rule \(\sigma \times \varepsilon = \sigma\) is simply a consequence of the fact that Wilson
lines can terminate on an anyon without creating flux, while
\(\varepsilon \times \varepsilon = 1\) is an immediate consequence of the definition of \(\varepsilon\).

\para\emph{Gauge-invariant Majorana swaps --} Since \(L\)-type links pair
Majoranas, edge rearrangements in the graph correspond to Majorana
swaps. Naively, to ``move'' Majoranas from position 1 to position 2, i.e.,
\(\alpha_1 \to \pm \alpha_2\) with some unitary \(\tilde{U}_{\pm}\), we mean
\(\tilde{U}_{\pm}^{\dagger} \alpha_2 \tilde{U}_{\pm} = \pm \alpha_1\). If
\(\alpha_1, \alpha_2\) are on different Majorana diamonds \(q, q'\), such a
\(\tilde{U}_{\pm}\) cannot be gauge invariant. The reason is that
\(\{\Gamma_q, \alpha_1\} = 0\), but \([\Gamma_q, \alpha_2] = 0\), so necessarily
\([\Gamma_q, \tilde{U}_{\pm}] \ne 0\).  In other words, \(\tilde{U}_{\pm}\) takes the
state away from the gauge-invariant Hilbert space. The simplest non-gauge
invariant swap is
\(\tilde{U}_{\pm} = \exp \left( \pm \frac{\pi}{4} \alpha_2 \alpha_1 \right)\), which also
takes \(\alpha_2 \to \mp \alpha_1\). The closest gauge invariant operator requires a path
\(\gamma\) from \(\alpha_1 \to \alpha_2\), from which we define,
\begin{equation}
  \label{eq:gauge-invariant-swap}
  U_{\pm} = \exp \left( \mp i \frac{\pi}{4} \mathcal{W}_{\gamma} \right).
\end{equation}
For this particular unitary, we can see explicitly how Wilson lines are
extended as the Majoranas are swapped
\begin{equation} 
  U_{\pm}^{\dagger}(\gamma) \alpha_2 W_{\gamma} U_{\pm}(\gamma) = \pm \alpha_1, \; U_{\pm}^{\dagger}(\gamma) \alpha_1 U_{\pm}(\gamma) = \mp \alpha_2 W_{\gamma}.
\end{equation}
We note that as long as a PSC is chosen where all the \(\tau_l\) are Pauli
operators, \(U_{\pm}\) is always in the Clifford group, and can therefore be
decomposed efficiently to \(\operatorname{CNOT}\) (or \(\operatorname{CZ}\))
and single-qubit Clifford gates.

\para \emph{Gates for moving anyons --} Graphically, moving a single anyon
from vertex 1 to vertex 2 corresponds to a rearrangment of \(L\)-type links
{[See Figs.~4(d-f)]}. The corresponding swaps \Cref{eq:gauge-invariant-swap}
are built from paths \(\gamma\) that run between an anyon \(\alpha_1\) and a Majorana
\(\alpha_2\) paired by an \(L\)-type edge to \(\alpha_3\) {[See Figs.~4 (d-f)]}. To
ensure the graph remains locally planar, it is sufficient to build larger
moves from elements where \(\alpha_1\) and \(\alpha_2\) share the same stabilizer
plaquette \(P\). There is a unique allowed path \(\gamma\) between them within
\(P\). Similarly to the line for \(B(P)\), in general this path is not
canonical. The sign in \Cref{eq:gauge-invariant-swap} is determined by the
condition that no flux is created in the new graph (with an \(L\)-type edge
between 1 and 3). Specifically, if the path \(\gamma\) is counter-clockwise about
the plaquette containing \(\alpha_2\) and \(\alpha_1\), we use the \(U_-\) {[See
  Fig.~4(d)]}. If the path is clockwise about this plaquette, we find
\(U_\zeta\) with \(\zeta=(-1)^{N_{ll}(\gamma)}\) where \(N_{ll}(\gamma)\) is the number of
vertices with adjacent \(l\)-type edges in \(\gamma\), defined in
\Cref{eq:aW-pauli} {[see Figs.~4(e) and (f)]}. By construction, this is an
example of a unitary motion of anyons without creation of
\(\mathbb{Z}_2^{(s)}\) flux. It follows that the canonicity of Wilson lines connecting
anyons is always preserved, despite the fact we chose to use a non-canonical
line give the rules for the unitaries. Finally, we remark that to move the
composite of an anyon and \(\mathbb{Z}_2^{(s)}\) flux, one simply uses the opposite
sign in \(U_{\pm}\) to the one for the bare anyon.

\para \emph{Braid generators --} Figures 4(g-h) show one minimal
implementation of the fundamental generator of the braid group,
\(R_{23}\). All other generators can be constructed in an analogous
manner. One advantage of this protocol is that it restores the lattice:
practically, this means such generators can be iterated an arbitrary number
of times, and theoretically it allows directly comparing states before and
after braiding. Another advantage is that it can be implemented on small
systems, and simply extended to make use of larger ones. The version shown
requires only 10 qubits and can therefore be implemented on existing
devices~\cite{satzingerRealizingTopologicallyOrdered2021}. A direct
experiment to establish non-Abelian statistics is to perform the lattice
version of Fig. 3 (c): simply create two anyons from the vacuum at the
locations \(t=1\), and perform this braid twice to implement
\(R_{23}^2\). After \(R_{23}^2\) each pair of anyons will fuse to an \(\varepsilon\).

In the future, periodic measurements of stabilizers would allow quantum error
correction, with the distance between anyons serving as an effective code
distance. On a larger device, extending the protocol in Figs.~4(g-h) simply
by starting the anyons further apart, and continuing the vertical motion of
the initially rightmost anyon at \(t = 3\), would allow maintaining a larger
code distance\footnote{More complicated braids can achieve larger code
  distances by constant factors in certain cases.}. The protocol involves
local code deformations, as a result of which the graph and stabilizer sizes
change, but the most non-local stabilizers can be restricted to be the
smallest possible 5-local operators\footnote{The precise procedure to
  accomplish this depends slightly on the available geometry when the devices
  are small, for example in many cases it is convenient to modify the step
  from \(t=2\) to \(t=3\) by moving edges to the \emph{right} instead of the
  left of the anyon to move it upwards. Such an extension is shown in
  Appendix A.}. We leave the analysis of this overhead to future work.

\para \emph{A GHZ experiment --} Another key element of topological quantum
computation is preparation of an entangled state of anyons. We give a
protocol such that a single braid takes a logical product state \(\ket{000}\)
to a GHZ state\cite{greenbergerGoingBellTheorem2007}, which is a starting
point for the discussion of multi-qubit entanglement. Our protocol also
serves as a concise demonstration of computational primitives introduced
above. Observe that the standard surface code encoding one logical qubit is
nothing else than our model with 4 Ising anyons at the
corners~\cite{bravyiCorrectingCoherentErrors2018}. We define logical $Z$
operators using the shortest Wilson lines for bulk anyons and an 't Hooft
line for the anyons at the corners, {[See Fig.~4(i)]}. The 't Hooft line is
chosen to run down the center of the sample, so that anyon pairs can be on
either side. When it splits anyon pairs, such an 't Hooft line is shorter
than any equivalent Wilson line.

To prepare the logical state \(\ket{000}\), it is simplest to start from the
\(\ket{0}\) state of the surface code, and create anyon pairs from the vacuum
at the locations shown in {Fig.~4(i)}. An exchange of bulk anyons 1 and 2
then prepares a state of the form
\(\ket{\text{GHZ}_{\phi}} = \frac{1}{\sqrt{2}} (\ket{000} + e^{i \phi}
\ket{111})\), where \(\phi\) depends on the phase choice of the logical
basis. To fix an unambiguous convention for \(\phi\) and perform full
tomography, it is sufficient to define logical \(X\) operators as in {Fig.~4(j)}. Then exchange of anyons 1 and 2 prepares
\(\ket{\text{GHZ}_{\pi/2}}\). An exchange of anyons 1 and 3, which can be
generated by conjugating the above braid with an exchange of 2 and 3,
prepares \(\ket{\text{GHZ}_{0}}\).

\para To summarize, we constructed a graph gauge theory with projective Ising
anyons. The consistency of the theory requires identification of two gauge
fields: one associated with the flux created by a plaquette (stabilizer)
violation and the other, the Kasteleyn orientation, is associated with the
flux carried by a D3V, degree three vertex of the graph. The presence of both
fields ensures that a loop physical path of an unpaired Majorana fermion
measures the number of unpaired Majoranas enclosed by it, giving rise to
non-Abelian braiding statistics. The formulation of physical operators in
terms of augmented Wilson lines and the graphical rules to construct them
allows a simple way to design unitary protocols for manipulation and
measurement of anyons. The unitary evolution can be thought of as the motion
of anyons directly realizing elementary braiding operations. We propose
specific experiments to realize the dynamics of anyons and verify their
fusion rules and braiding statistics as well as preparation of an entangled
state of anyons. The protocols we proposed were implemented experimentally on
a superconducting processor as reported in the forthcoming publication. Our
recipe for constructing protocols could be used to realize quantum
computation with non-Abelian anyons that allows for quantum error
correction.~\footnote{In our system of Ising anyons Clifford gates can be
  implemented fault-tolerantly. A non-Clifford $T$-gate necessary for
  universal computation can be constructed by replacing $\pi/4\to \pi/8$ in
  Eq.~(\ref{eq:gauge-invariant-swap}) and taking the line between any two
  anyons. This operation is not fault-tolerant.}

\noindent{\bf Acknowlegements:}
YL gratefully acknowledges conversations with Felipe Hernandez and Chao-Ming
Jian. We also thank Eduardo Fradkin, Bert Halperin, Trond Andersen, Pedram Roushan and members of Quantum AI for discussions. YL and EAK acknowledges support by a New Frontier Grant from Cornell
University’s College of Arts and Sciences. EAK acknowledges support by the
NSF under OAC-2118310, the Ewha Frontier 10-10 Research Grant, and the Simons
Fellowship in Theoretical Physics award 920665. EAK performed a part of this
work at the Aspen Center for Physics, which is supported by the National
Science Foundation grant PHY-160761.

\bibliography{bibliography.bib}

\begin{thebibliography}{48}%
\makeatletter
\providecommand \@ifxundefined [1]{%
 \@ifx{#1\undefined}
}%
\providecommand \@ifnum [1]{%
 \ifnum #1\expandafter \@firstoftwo
 \else \expandafter \@secondoftwo
 \fi
}%
\providecommand \@ifx [1]{%
 \ifx #1\expandafter \@firstoftwo
 \else \expandafter \@secondoftwo
 \fi
}%
\providecommand \natexlab [1]{#1}%
\providecommand \enquote  [1]{``#1''}%
\providecommand \bibnamefont  [1]{#1}%
\providecommand \bibfnamefont [1]{#1}%
\providecommand \citenamefont [1]{#1}%
\providecommand \href@noop [0]{\@secondoftwo}%
\providecommand \href [0]{\begingroup \@sanitize@url \@href}%
\providecommand \@href[1]{\@@startlink{#1}\@@href}%
\providecommand \@@href[1]{\endgroup#1\@@endlink}%
\providecommand \@sanitize@url [0]{\catcode `\\12\catcode `\$12\catcode
  `\&12\catcode `\#12\catcode `\^12\catcode `\_12\catcode `\%12\relax}%
\providecommand \@@startlink[1]{}%
\providecommand \@@endlink[0]{}%
\providecommand \url  [0]{\begingroup\@sanitize@url \@url }%
\providecommand \@url [1]{\endgroup\@href {#1}{\urlprefix }}%
\providecommand \urlprefix  [0]{URL }%
\providecommand \Eprint [0]{\href }%
\providecommand \doibase [0]{https://doi.org/}%
\providecommand \selectlanguage [0]{\@gobble}%
\providecommand \bibinfo  [0]{\@secondoftwo}%
\providecommand \bibfield  [0]{\@secondoftwo}%
\providecommand \translation [1]{[#1]}%
\providecommand \BibitemOpen [0]{}%
\providecommand \bibitemStop [0]{}%
\providecommand \bibitemNoStop [0]{.\EOS\space}%
\providecommand \EOS [0]{\spacefactor3000\relax}%
\providecommand \BibitemShut  [1]{\csname bibitem#1\endcsname}%
\let\auto@bib@innerbib\@empty
\bibitem [{\citenamefont
  {Kitaev}(2003)}]{kitaevFaulttolerantQuantumComputation2003}%
  \BibitemOpen
  \bibfield  {author} {\bibinfo {author} {\bibfnamefont {A.~Y.}\ \bibnamefont
  {Kitaev}},\ }\bibfield  {title} {\bibinfo {title} {Fault-tolerant quantum
  computation by anyons},\ }\href
  {https://doi.org/10.1016/S0003-4916(02)00018-0} {\bibfield  {journal}
  {\bibinfo  {journal} {Annals of Physics}\ }\textbf {\bibinfo {volume}
  {303}},\ \bibinfo {pages} {2} (\bibinfo {year} {2003})},\ \Eprint
  {https://arxiv.org/abs/quant-ph/9707021} {arXiv:quant-ph/9707021}
  \BibitemShut {NoStop}%
\bibitem [{\citenamefont {Nayak}\ \emph {et~al.}(2008)\citenamefont {Nayak},
  \citenamefont {Simon}, \citenamefont {Stern}, \citenamefont {Freedman},\ and\
  \citenamefont {Sarma}}]{nayakNonAbelianAnyonsTopological2008}%
  \BibitemOpen
  \bibfield  {author} {\bibinfo {author} {\bibfnamefont {C.}~\bibnamefont
  {Nayak}}, \bibinfo {author} {\bibfnamefont {S.~H.}\ \bibnamefont {Simon}},
  \bibinfo {author} {\bibfnamefont {A.}~\bibnamefont {Stern}}, \bibinfo
  {author} {\bibfnamefont {M.}~\bibnamefont {Freedman}},\ and\ \bibinfo
  {author} {\bibfnamefont {S.~D.}\ \bibnamefont {Sarma}},\ }\bibfield  {title}
  {\bibinfo {title} {Non-{{Abelian Anyons}} and {{Topological Quantum
  Computation}}},\ }\href {https://doi.org/10.1103/RevModPhys.80.1083}
  {\bibfield  {journal} {\bibinfo  {journal} {Reviews of Modern Physics}\
  }\textbf {\bibinfo {volume} {80}},\ \bibinfo {pages} {1083} (\bibinfo {year}
  {2008})},\ \Eprint {https://arxiv.org/abs/0707.1889} {arXiv:0707.1889}
  \BibitemShut {NoStop}%
\bibitem [{\citenamefont {Stern}(2010)}]{sternNonAbelianStatesMatter2010}%
  \BibitemOpen
  \bibfield  {author} {\bibinfo {author} {\bibfnamefont {A.}~\bibnamefont
  {Stern}},\ }\bibfield  {title} {\bibinfo {title} {Non-{{Abelian}} states of
  matter},\ }\href {https://doi.org/10.1038/nature08915} {\bibfield  {journal}
  {\bibinfo  {journal} {Nature}\ }\textbf {\bibinfo {volume} {464}},\ \bibinfo
  {pages} {187} (\bibinfo {year} {2010})}\BibitemShut {NoStop}%
\bibitem [{\citenamefont {Moore}\ and\ \citenamefont
  {Read}(1991)}]{mooreNonabelionsFractionalQuantum1991a}%
  \BibitemOpen
  \bibfield  {author} {\bibinfo {author} {\bibfnamefont {G.}~\bibnamefont
  {Moore}}\ and\ \bibinfo {author} {\bibfnamefont {N.}~\bibnamefont {Read}},\
  }\bibfield  {title} {\bibinfo {title} {Nonabelions in the fractional quantum
  hall effect},\ }\href {https://doi.org/10.1016/0550-3213(91)90407-O}
  {\bibfield  {journal} {\bibinfo  {journal} {Nuclear Physics B}\ }\textbf
  {\bibinfo {volume} {360}},\ \bibinfo {pages} {362} (\bibinfo {year}
  {1991})}\BibitemShut {NoStop}%
\bibitem [{\citenamefont {Nayak}\ and\ \citenamefont
  {Wilczek}(1996)}]{nayak2nquasiholeStatesRealize1996}%
  \BibitemOpen
  \bibfield  {author} {\bibinfo {author} {\bibfnamefont {C.}~\bibnamefont
  {Nayak}}\ and\ \bibinfo {author} {\bibfnamefont {F.}~\bibnamefont
  {Wilczek}},\ }\bibfield  {title} {\bibinfo {title} {2n-quasihole states
  realize 2n-1-dimensional spinor braiding statistics in paired quantum
  {{Hall}} states},\ }\href {https://doi.org/10.1016/0550-3213(96)00430-0}
  {\bibfield  {journal} {\bibinfo  {journal} {Nuclear Physics B}\ }\textbf
  {\bibinfo {volume} {479}},\ \bibinfo {pages} {529} (\bibinfo {year}
  {1996})}\BibitemShut {NoStop}%
\bibitem [{\citenamefont {Kitaev}(2006)}]{kitaevAnyonsExactlySolved2006}%
  \BibitemOpen
  \bibfield  {author} {\bibinfo {author} {\bibfnamefont {A.}~\bibnamefont
  {Kitaev}},\ }\bibfield  {title} {\bibinfo {title} {Anyons in an exactly
  solved model and beyond},\ }\href {https://doi.org/10.1016/j.aop.2005.10.005}
  {\bibfield  {journal} {\bibinfo  {journal} {Annals of Physics}\ }\textbf
  {\bibinfo {volume} {321}},\ \bibinfo {pages} {2} (\bibinfo {year} {2006})},\
  \Eprint {https://arxiv.org/abs/cond-mat/0506438} {arXiv:cond-mat/0506438}
  \BibitemShut {NoStop}%
\bibitem [{\citenamefont
  {Ivanov}(2001)}]{ivanovNonabelianStatisticsHalfquantum2001}%
  \BibitemOpen
  \bibfield  {author} {\bibinfo {author} {\bibfnamefont {D.~A.}\ \bibnamefont
  {Ivanov}},\ }\bibfield  {title} {\bibinfo {title} {Non-abelian statistics of
  half-quantum vortices in p-wave superconductors},\ }\href
  {https://doi.org/10.1103/PhysRevLett.86.268} {\bibfield  {journal} {\bibinfo
  {journal} {Physical Review Letters}\ }\textbf {\bibinfo {volume} {86}},\
  \bibinfo {pages} {268} (\bibinfo {year} {2001})},\ \Eprint
  {https://arxiv.org/abs/cond-mat/0005069} {arXiv:cond-mat/0005069}
  \BibitemShut {NoStop}%
\bibitem [{\citenamefont {Read}\ and\ \citenamefont
  {Green}(2000)}]{readPairedStatesFermions2000a}%
  \BibitemOpen
  \bibfield  {author} {\bibinfo {author} {\bibfnamefont {N.}~\bibnamefont
  {Read}}\ and\ \bibinfo {author} {\bibfnamefont {D.}~\bibnamefont {Green}},\
  }\bibfield  {title} {\bibinfo {title} {Paired states of fermions in two
  dimensions with breaking of parity and time-reversal symmetries, and the
  fractional quantum {{Hall}} effect},\ }\href
  {https://doi.org/10.1103/PhysRevB.61.10267} {\bibfield  {journal} {\bibinfo
  {journal} {Physical Review B}\ }\textbf {\bibinfo {volume} {61}},\ \bibinfo
  {pages} {10267} (\bibinfo {year} {2000})},\ \Eprint
  {https://arxiv.org/abs/cond-mat/9906453} {arXiv:cond-mat/9906453}
  \BibitemShut {NoStop}%
\bibitem [{\citenamefont {Bonderson}\ \emph {et~al.}(2006)\citenamefont
  {Bonderson}, \citenamefont {Kitaev},\ and\ \citenamefont
  {Shtengel}}]{bondersonDetectingNonAbelianStatistics2006}%
  \BibitemOpen
  \bibfield  {author} {\bibinfo {author} {\bibfnamefont {P.}~\bibnamefont
  {Bonderson}}, \bibinfo {author} {\bibfnamefont {A.}~\bibnamefont {Kitaev}},\
  and\ \bibinfo {author} {\bibfnamefont {K.}~\bibnamefont {Shtengel}},\
  }\bibfield  {title} {\bibinfo {title} {Detecting {{Non-Abelian Statistics}}
  in the nu=5/2 {{Fractional Quantum Hall State}}},\ }\href
  {https://doi.org/10.1103/PhysRevLett.96.016803} {\bibfield  {journal}
  {\bibinfo  {journal} {Physical Review Letters}\ }\textbf {\bibinfo {volume}
  {96}},\ \bibinfo {pages} {016803} (\bibinfo {year} {2006})}\BibitemShut
  {NoStop}%
\bibitem [{\citenamefont {Clarke}\ \emph {et~al.}(2011)\citenamefont {Clarke},
  \citenamefont {Sau},\ and\ \citenamefont
  {Tewari}}]{clarkeMajoranaFermionExchange2011a}%
  \BibitemOpen
  \bibfield  {author} {\bibinfo {author} {\bibfnamefont {D.~J.}\ \bibnamefont
  {Clarke}}, \bibinfo {author} {\bibfnamefont {J.~D.}\ \bibnamefont {Sau}},\
  and\ \bibinfo {author} {\bibfnamefont {S.}~\bibnamefont {Tewari}},\
  }\bibfield  {title} {\bibinfo {title} {Majorana fermion exchange in
  quasi-one-dimensional networks},\ }\href
  {https://doi.org/10.1103/PhysRevB.84.035120} {\bibfield  {journal} {\bibinfo
  {journal} {Physical Review B}\ }\textbf {\bibinfo {volume} {84}},\ \bibinfo
  {pages} {035120} (\bibinfo {year} {2011})},\ \Eprint
  {https://arxiv.org/abs/1012.0296} {arXiv:1012.0296 [cond-mat]} \BibitemShut
  {NoStop}%
\bibitem [{\citenamefont {Alicea}\ \emph {et~al.}(2011)\citenamefont {Alicea},
  \citenamefont {Oreg}, \citenamefont {Refael}, \citenamefont {{von Oppen}},\
  and\ \citenamefont {Fisher}}]{aliceaNonAbelianStatisticsTopological2011b}%
  \BibitemOpen
  \bibfield  {author} {\bibinfo {author} {\bibfnamefont {J.}~\bibnamefont
  {Alicea}}, \bibinfo {author} {\bibfnamefont {Y.}~\bibnamefont {Oreg}},
  \bibinfo {author} {\bibfnamefont {G.}~\bibnamefont {Refael}}, \bibinfo
  {author} {\bibfnamefont {F.}~\bibnamefont {{von Oppen}}},\ and\ \bibinfo
  {author} {\bibfnamefont {M.~P.~A.}\ \bibnamefont {Fisher}},\ }\bibfield
  {title} {\bibinfo {title} {Non-{{Abelian}} statistics and topological quantum
  information processing in {{1D}} wire networks},\ }\href
  {https://doi.org/10.1038/nphys1915} {\bibfield  {journal} {\bibinfo
  {journal} {Nature Physics}\ }\textbf {\bibinfo {volume} {7}},\ \bibinfo
  {pages} {412} (\bibinfo {year} {2011})},\ \Eprint
  {https://arxiv.org/abs/1006.4395} {arXiv:1006.4395 [cond-mat]} \BibitemShut
  {NoStop}%
\bibitem [{\citenamefont {Lindner}\ \emph {et~al.}(2012)\citenamefont
  {Lindner}, \citenamefont {Berg}, \citenamefont {Refael},\ and\ \citenamefont
  {Stern}}]{lindnerFractionalizingMajoranaFermions2012a}%
  \BibitemOpen
  \bibfield  {author} {\bibinfo {author} {\bibfnamefont {N.~H.}\ \bibnamefont
  {Lindner}}, \bibinfo {author} {\bibfnamefont {E.}~\bibnamefont {Berg}},
  \bibinfo {author} {\bibfnamefont {G.}~\bibnamefont {Refael}},\ and\ \bibinfo
  {author} {\bibfnamefont {A.}~\bibnamefont {Stern}},\ }\bibfield  {title}
  {\bibinfo {title} {Fractionalizing {{Majorana}} fermions: Non-abelian
  statistics on the edges of abelian quantum {{Hall}} states},\ }\href
  {https://doi.org/10.1103/PhysRevX.2.041002} {\bibfield  {journal} {\bibinfo
  {journal} {Physical Review X}\ }\textbf {\bibinfo {volume} {2}},\ \bibinfo
  {pages} {041002} (\bibinfo {year} {2012})},\ \Eprint
  {https://arxiv.org/abs/1204.5733} {arXiv:1204.5733 [cond-mat]} \BibitemShut
  {NoStop}%
\bibitem [{\citenamefont {Stern}\ and\ \citenamefont
  {Lindner}(2013)}]{sternTopologicalQuantumComputation2013a}%
  \BibitemOpen
  \bibfield  {author} {\bibinfo {author} {\bibfnamefont {A.}~\bibnamefont
  {Stern}}\ and\ \bibinfo {author} {\bibfnamefont {N.~H.}\ \bibnamefont
  {Lindner}},\ }\bibfield  {title} {\bibinfo {title} {Topological {{Quantum
  Computation}}\textemdash{{From Basic Concepts}} to {{First Experiments}}},\
  }\href {https://doi.org/10.1126/science.1231473} {\bibfield  {journal}
  {\bibinfo  {journal} {Science}\ }\textbf {\bibinfo {volume} {339}},\ \bibinfo
  {pages} {1179} (\bibinfo {year} {2013})}\BibitemShut {NoStop}%
\bibitem [{\citenamefont {Clarke}\ \emph {et~al.}(2013)\citenamefont {Clarke},
  \citenamefont {Alicea},\ and\ \citenamefont
  {Shtengel}}]{clarkeExoticNonAbelianAnyons2013}%
  \BibitemOpen
  \bibfield  {author} {\bibinfo {author} {\bibfnamefont {D.~J.}\ \bibnamefont
  {Clarke}}, \bibinfo {author} {\bibfnamefont {J.}~\bibnamefont {Alicea}},\
  and\ \bibinfo {author} {\bibfnamefont {K.}~\bibnamefont {Shtengel}},\
  }\bibfield  {title} {\bibinfo {title} {Exotic non-{{Abelian}} anyons from
  conventional fractional quantum {{Hall}} states},\ }\href
  {https://doi.org/10.1038/ncomms2340} {\bibfield  {journal} {\bibinfo
  {journal} {Nature Communications}\ }\textbf {\bibinfo {volume} {4}},\
  \bibinfo {pages} {1348} (\bibinfo {year} {2013})},\ \Eprint
  {https://arxiv.org/abs/1204.5479} {arXiv:1204.5479 [cond-mat]} \BibitemShut
  {NoStop}%
\bibitem [{\citenamefont {Satzinger}\ \emph {et~al.}(2021)\citenamefont
  {Satzinger}, \citenamefont {Liu}, \citenamefont {Smith}, \citenamefont
  {Knapp}, \citenamefont {Newman}, \citenamefont {Jones}, \citenamefont {Chen},
  \citenamefont {Quintana}, \citenamefont {Mi}, \citenamefont {Dunsworth},
  \citenamefont {Gidney}, \citenamefont {Aleiner}, \citenamefont {Arute},
  \citenamefont {Arya}, \citenamefont {Atalaya}, \citenamefont {Babbush},
  \citenamefont {Bardin}, \citenamefont {Barends}, \citenamefont {Basso},
  \citenamefont {Bengtsson}, \citenamefont {Bilmes}, \citenamefont {Broughton},
  \citenamefont {Buckley}, \citenamefont {Buell}, \citenamefont {Burkett},
  \citenamefont {Bushnell}, \citenamefont {Chiaro}, \citenamefont {Collins},
  \citenamefont {Courtney}, \citenamefont {Demura}, \citenamefont {Derk},
  \citenamefont {Eppens}, \citenamefont {Erickson}, \citenamefont {Farhi},
  \citenamefont {Foaro}, \citenamefont {Fowler}, \citenamefont {Foxen},
  \citenamefont {Giustina}, \citenamefont {Greene}, \citenamefont {Gross},
  \citenamefont {Harrigan}, \citenamefont {Harrington}, \citenamefont {Hilton},
  \citenamefont {Hong}, \citenamefont {Huang}, \citenamefont {Huggins},
  \citenamefont {Ioffe}, \citenamefont {Isakov}, \citenamefont {Jeffrey},
  \citenamefont {Jiang}, \citenamefont {Kafri}, \citenamefont {Kechedzhi},
  \citenamefont {Khattar}, \citenamefont {Kim}, \citenamefont {Klimov},
  \citenamefont {Korotkov}, \citenamefont {Kostritsa}, \citenamefont
  {Landhuis}, \citenamefont {Laptev}, \citenamefont {Locharla}, \citenamefont
  {Lucero}, \citenamefont {Martin}, \citenamefont {McClean}, \citenamefont
  {McEwen}, \citenamefont {Miao}, \citenamefont {Mohseni}, \citenamefont
  {Montazeri}, \citenamefont {Mruczkiewicz}, \citenamefont {Mutus},
  \citenamefont {Naaman}, \citenamefont {Neeley}, \citenamefont {Neill},
  \citenamefont {Niu}, \citenamefont {O'Brien}, \citenamefont {Opremcak},
  \citenamefont {Pat{\'o}}, \citenamefont {Petukhov}, \citenamefont {Rubin},
  \citenamefont {Sank}, \citenamefont {Shvarts}, \citenamefont {Strain},
  \citenamefont {Szalay}, \citenamefont {Villalonga}, \citenamefont {White},
  \citenamefont {Yao}, \citenamefont {Yeh}, \citenamefont {Yoo}, \citenamefont
  {Zalcman}, \citenamefont {Neven}, \citenamefont {Boixo}, \citenamefont
  {Megrant}, \citenamefont {Chen}, \citenamefont {Kelly}, \citenamefont
  {Smelyanskiy}, \citenamefont {Kitaev}, \citenamefont {Knap}, \citenamefont
  {Pollmann},\ and\ \citenamefont
  {Roushan}}]{satzingerRealizingTopologicallyOrdered2021}%
  \BibitemOpen
  \bibfield  {author} {\bibinfo {author} {\bibfnamefont {K.~J.}\ \bibnamefont
  {Satzinger}}, \bibinfo {author} {\bibfnamefont {Y.}~\bibnamefont {Liu}},
  \bibinfo {author} {\bibfnamefont {A.}~\bibnamefont {Smith}}, \bibinfo
  {author} {\bibfnamefont {C.}~\bibnamefont {Knapp}}, \bibinfo {author}
  {\bibfnamefont {M.}~\bibnamefont {Newman}}, \bibinfo {author} {\bibfnamefont
  {C.}~\bibnamefont {Jones}}, \bibinfo {author} {\bibfnamefont
  {Z.}~\bibnamefont {Chen}}, \bibinfo {author} {\bibfnamefont {C.}~\bibnamefont
  {Quintana}}, \bibinfo {author} {\bibfnamefont {X.}~\bibnamefont {Mi}},
  \bibinfo {author} {\bibfnamefont {A.}~\bibnamefont {Dunsworth}}, \bibinfo
  {author} {\bibfnamefont {C.}~\bibnamefont {Gidney}}, \bibinfo {author}
  {\bibfnamefont {I.}~\bibnamefont {Aleiner}}, \bibinfo {author} {\bibfnamefont
  {F.}~\bibnamefont {Arute}}, \bibinfo {author} {\bibfnamefont
  {K.}~\bibnamefont {Arya}}, \bibinfo {author} {\bibfnamefont {J.}~\bibnamefont
  {Atalaya}}, \bibinfo {author} {\bibfnamefont {R.}~\bibnamefont {Babbush}},
  \bibinfo {author} {\bibfnamefont {J.~C.}\ \bibnamefont {Bardin}}, \bibinfo
  {author} {\bibfnamefont {R.}~\bibnamefont {Barends}}, \bibinfo {author}
  {\bibfnamefont {J.}~\bibnamefont {Basso}}, \bibinfo {author} {\bibfnamefont
  {A.}~\bibnamefont {Bengtsson}}, \bibinfo {author} {\bibfnamefont
  {A.}~\bibnamefont {Bilmes}}, \bibinfo {author} {\bibfnamefont
  {M.}~\bibnamefont {Broughton}}, \bibinfo {author} {\bibfnamefont {B.~B.}\
  \bibnamefont {Buckley}}, \bibinfo {author} {\bibfnamefont {D.~A.}\
  \bibnamefont {Buell}}, \bibinfo {author} {\bibfnamefont {B.}~\bibnamefont
  {Burkett}}, \bibinfo {author} {\bibfnamefont {N.}~\bibnamefont {Bushnell}},
  \bibinfo {author} {\bibfnamefont {B.}~\bibnamefont {Chiaro}}, \bibinfo
  {author} {\bibfnamefont {R.}~\bibnamefont {Collins}}, \bibinfo {author}
  {\bibfnamefont {W.}~\bibnamefont {Courtney}}, \bibinfo {author}
  {\bibfnamefont {S.}~\bibnamefont {Demura}}, \bibinfo {author} {\bibfnamefont
  {A.~R.}\ \bibnamefont {Derk}}, \bibinfo {author} {\bibfnamefont
  {D.}~\bibnamefont {Eppens}}, \bibinfo {author} {\bibfnamefont
  {C.}~\bibnamefont {Erickson}}, \bibinfo {author} {\bibfnamefont
  {E.}~\bibnamefont {Farhi}}, \bibinfo {author} {\bibfnamefont
  {L.}~\bibnamefont {Foaro}}, \bibinfo {author} {\bibfnamefont {A.~G.}\
  \bibnamefont {Fowler}}, \bibinfo {author} {\bibfnamefont {B.}~\bibnamefont
  {Foxen}}, \bibinfo {author} {\bibfnamefont {M.}~\bibnamefont {Giustina}},
  \bibinfo {author} {\bibfnamefont {A.}~\bibnamefont {Greene}}, \bibinfo
  {author} {\bibfnamefont {J.~A.}\ \bibnamefont {Gross}}, \bibinfo {author}
  {\bibfnamefont {M.~P.}\ \bibnamefont {Harrigan}}, \bibinfo {author}
  {\bibfnamefont {S.~D.}\ \bibnamefont {Harrington}}, \bibinfo {author}
  {\bibfnamefont {J.}~\bibnamefont {Hilton}}, \bibinfo {author} {\bibfnamefont
  {S.}~\bibnamefont {Hong}}, \bibinfo {author} {\bibfnamefont {T.}~\bibnamefont
  {Huang}}, \bibinfo {author} {\bibfnamefont {W.~J.}\ \bibnamefont {Huggins}},
  \bibinfo {author} {\bibfnamefont {L.~B.}\ \bibnamefont {Ioffe}}, \bibinfo
  {author} {\bibfnamefont {S.~V.}\ \bibnamefont {Isakov}}, \bibinfo {author}
  {\bibfnamefont {E.}~\bibnamefont {Jeffrey}}, \bibinfo {author} {\bibfnamefont
  {Z.}~\bibnamefont {Jiang}}, \bibinfo {author} {\bibfnamefont
  {D.}~\bibnamefont {Kafri}}, \bibinfo {author} {\bibfnamefont
  {K.}~\bibnamefont {Kechedzhi}}, \bibinfo {author} {\bibfnamefont
  {T.}~\bibnamefont {Khattar}}, \bibinfo {author} {\bibfnamefont
  {S.}~\bibnamefont {Kim}}, \bibinfo {author} {\bibfnamefont {P.~V.}\
  \bibnamefont {Klimov}}, \bibinfo {author} {\bibfnamefont {A.~N.}\
  \bibnamefont {Korotkov}}, \bibinfo {author} {\bibfnamefont {F.}~\bibnamefont
  {Kostritsa}}, \bibinfo {author} {\bibfnamefont {D.}~\bibnamefont {Landhuis}},
  \bibinfo {author} {\bibfnamefont {P.}~\bibnamefont {Laptev}}, \bibinfo
  {author} {\bibfnamefont {A.}~\bibnamefont {Locharla}}, \bibinfo {author}
  {\bibfnamefont {E.}~\bibnamefont {Lucero}}, \bibinfo {author} {\bibfnamefont
  {O.}~\bibnamefont {Martin}}, \bibinfo {author} {\bibfnamefont {J.~R.}\
  \bibnamefont {McClean}}, \bibinfo {author} {\bibfnamefont {M.}~\bibnamefont
  {McEwen}}, \bibinfo {author} {\bibfnamefont {K.~C.}\ \bibnamefont {Miao}},
  \bibinfo {author} {\bibfnamefont {M.}~\bibnamefont {Mohseni}}, \bibinfo
  {author} {\bibfnamefont {S.}~\bibnamefont {Montazeri}}, \bibinfo {author}
  {\bibfnamefont {W.}~\bibnamefont {Mruczkiewicz}}, \bibinfo {author}
  {\bibfnamefont {J.}~\bibnamefont {Mutus}}, \bibinfo {author} {\bibfnamefont
  {O.}~\bibnamefont {Naaman}}, \bibinfo {author} {\bibfnamefont
  {M.}~\bibnamefont {Neeley}}, \bibinfo {author} {\bibfnamefont
  {C.}~\bibnamefont {Neill}}, \bibinfo {author} {\bibfnamefont {M.~Y.}\
  \bibnamefont {Niu}}, \bibinfo {author} {\bibfnamefont {T.~E.}\ \bibnamefont
  {O'Brien}}, \bibinfo {author} {\bibfnamefont {A.}~\bibnamefont {Opremcak}},
  \bibinfo {author} {\bibfnamefont {B.}~\bibnamefont {Pat{\'o}}}, \bibinfo
  {author} {\bibfnamefont {A.}~\bibnamefont {Petukhov}}, \bibinfo {author}
  {\bibfnamefont {N.~C.}\ \bibnamefont {Rubin}}, \bibinfo {author}
  {\bibfnamefont {D.}~\bibnamefont {Sank}}, \bibinfo {author} {\bibfnamefont
  {V.}~\bibnamefont {Shvarts}}, \bibinfo {author} {\bibfnamefont
  {D.}~\bibnamefont {Strain}}, \bibinfo {author} {\bibfnamefont
  {M.}~\bibnamefont {Szalay}}, \bibinfo {author} {\bibfnamefont
  {B.}~\bibnamefont {Villalonga}}, \bibinfo {author} {\bibfnamefont {T.~C.}\
  \bibnamefont {White}}, \bibinfo {author} {\bibfnamefont {Z.}~\bibnamefont
  {Yao}}, \bibinfo {author} {\bibfnamefont {P.}~\bibnamefont {Yeh}}, \bibinfo
  {author} {\bibfnamefont {J.}~\bibnamefont {Yoo}}, \bibinfo {author}
  {\bibfnamefont {A.}~\bibnamefont {Zalcman}}, \bibinfo {author} {\bibfnamefont
  {H.}~\bibnamefont {Neven}}, \bibinfo {author} {\bibfnamefont
  {S.}~\bibnamefont {Boixo}}, \bibinfo {author} {\bibfnamefont
  {A.}~\bibnamefont {Megrant}}, \bibinfo {author} {\bibfnamefont
  {Y.}~\bibnamefont {Chen}}, \bibinfo {author} {\bibfnamefont {J.}~\bibnamefont
  {Kelly}}, \bibinfo {author} {\bibfnamefont {V.}~\bibnamefont {Smelyanskiy}},
  \bibinfo {author} {\bibfnamefont {A.}~\bibnamefont {Kitaev}}, \bibinfo
  {author} {\bibfnamefont {M.}~\bibnamefont {Knap}}, \bibinfo {author}
  {\bibfnamefont {F.}~\bibnamefont {Pollmann}},\ and\ \bibinfo {author}
  {\bibfnamefont {P.}~\bibnamefont {Roushan}},\ }\bibfield  {title} {\bibinfo
  {title} {Realizing topologically ordered states on a quantum processor},\
  }\href {https://doi.org/10.1126/science.abi8378} {\bibfield  {journal}
  {\bibinfo  {journal} {Science}\ }\textbf {\bibinfo {volume} {374}},\ \bibinfo
  {pages} {1237} (\bibinfo {year} {2021})},\ \Eprint
  {https://arxiv.org/abs/2104.01180} {arXiv:2104.01180} \BibitemShut {NoStop}%
\bibitem [{\citenamefont
  {Gottesman}(1997)}]{gottesmanStabilizerCodesQuantum1997}%
  \BibitemOpen
  \bibfield  {author} {\bibinfo {author} {\bibfnamefont {D.}~\bibnamefont
  {Gottesman}},\ }\bibfield  {title} {\bibinfo {title} {Stabilizer {{Codes}}
  and {{Quantum Error Correction}}},\ }\href@noop {} {\bibfield  {journal}
  {\bibinfo  {journal} {arXiv:quant-ph/9705052}\ } (\bibinfo {year} {1997})},\
  \Eprint {https://arxiv.org/abs/quant-ph/9705052} {arXiv:quant-ph/9705052}
  \BibitemShut {NoStop}%
\bibitem [{\citenamefont {Wen}(2003)}]{wenQuantumOrdersExact2003a}%
  \BibitemOpen
  \bibfield  {author} {\bibinfo {author} {\bibfnamefont {X.-G.}\ \bibnamefont
  {Wen}},\ }\bibfield  {title} {\bibinfo {title} {Quantum orders in an exact
  soluble model},\ }\href {https://doi.org/10.1103/PhysRevLett.90.016803}
  {\bibfield  {journal} {\bibinfo  {journal} {Physical Review Letters}\
  }\textbf {\bibinfo {volume} {90}},\ \bibinfo {pages} {016803} (\bibinfo
  {year} {2003})},\ \Eprint {https://arxiv.org/abs/quant-ph/0205004}
  {arXiv:quant-ph/0205004} \BibitemShut {NoStop}%
\bibitem [{\citenamefont {Bravyi}\ and\ \citenamefont
  {Kitaev}(1998)}]{bravyiQuantumCodesLattice1998}%
  \BibitemOpen
  \bibfield  {author} {\bibinfo {author} {\bibfnamefont {S.~B.}\ \bibnamefont
  {Bravyi}}\ and\ \bibinfo {author} {\bibfnamefont {A.~Y.}\ \bibnamefont
  {Kitaev}},\ }\bibfield  {title} {\bibinfo {title} {Quantum codes on a lattice
  with boundary},\ }\href@noop {} {\bibfield  {journal} {\bibinfo  {journal}
  {arXiv:quant-ph/9811052}\ } (\bibinfo {year} {1998})},\ \Eprint
  {https://arxiv.org/abs/quant-ph/9811052} {arXiv:quant-ph/9811052}
  \BibitemShut {NoStop}%
\bibitem [{\citenamefont {Freedman}\ and\ \citenamefont
  {Meyer}(1998)}]{freedmanProjectivePlanePlanar1998}%
  \BibitemOpen
  \bibfield  {author} {\bibinfo {author} {\bibfnamefont {M.~H.}\ \bibnamefont
  {Freedman}}\ and\ \bibinfo {author} {\bibfnamefont {D.~A.}\ \bibnamefont
  {Meyer}},\ }\href {https://doi.org/10.48550/arXiv.quant-ph/9810055} {\bibinfo
  {title} {Projective plane and planar quantum codes}} (\bibinfo {year}
  {1998}),\ \Eprint {https://arxiv.org/abs/quant-ph/9810055}
  {arXiv:quant-ph/9810055} \BibitemShut {NoStop}%
\bibitem [{\citenamefont {Bombin}(2010)}]{bombinTopologicalOrderTwist2010}%
  \BibitemOpen
  \bibfield  {author} {\bibinfo {author} {\bibfnamefont {H.}~\bibnamefont
  {Bombin}},\ }\bibfield  {title} {\bibinfo {title} {Topological {{Order}} with
  a {{Twist}}: {{Ising Anyons}} from an {{Abelian Model}}},\ }\href
  {https://doi.org/10.1103/PhysRevLett.105.030403} {\bibfield  {journal}
  {\bibinfo  {journal} {Physical Review Letters}\ }\textbf {\bibinfo {volume}
  {105}},\ \bibinfo {pages} {030403} (\bibinfo {year} {2010})},\ \Eprint
  {https://arxiv.org/abs/1004.1838} {arXiv:1004.1838 [cond-mat, physics:hep-th,
  physics:quant-ph]} \BibitemShut {NoStop}%
\bibitem [{\citenamefont {Kitaev}\ and\ \citenamefont
  {Kong}(2012)}]{kitaevModelsGappedBoundaries2012a}%
  \BibitemOpen
  \bibfield  {author} {\bibinfo {author} {\bibfnamefont {A.}~\bibnamefont
  {Kitaev}}\ and\ \bibinfo {author} {\bibfnamefont {L.}~\bibnamefont {Kong}},\
  }\bibfield  {title} {\bibinfo {title} {Models for gapped boundaries and
  domain walls},\ }\href {https://doi.org/10.1007/s00220-012-1500-5} {\bibfield
   {journal} {\bibinfo  {journal} {Communications in Mathematical Physics}\
  }\textbf {\bibinfo {volume} {313}},\ \bibinfo {pages} {351} (\bibinfo {year}
  {2012})},\ \Eprint {https://arxiv.org/abs/1104.5047} {arXiv:1104.5047
  [cond-mat]} \BibitemShut {NoStop}%
\bibitem [{\citenamefont {Barkeshli}\ \emph {et~al.}(2013)\citenamefont
  {Barkeshli}, \citenamefont {Jian},\ and\ \citenamefont
  {Qi}}]{barkeshliTheoryDefectsAbelian2013}%
  \BibitemOpen
  \bibfield  {author} {\bibinfo {author} {\bibfnamefont {M.}~\bibnamefont
  {Barkeshli}}, \bibinfo {author} {\bibfnamefont {C.-M.}\ \bibnamefont
  {Jian}},\ and\ \bibinfo {author} {\bibfnamefont {X.-L.}\ \bibnamefont {Qi}},\
  }\bibfield  {title} {\bibinfo {title} {Theory of defects in {{Abelian}}
  topological states},\ }\href {https://doi.org/10.1103/PhysRevB.88.235103}
  {\bibfield  {journal} {\bibinfo  {journal} {Physical Review B}\ }\textbf
  {\bibinfo {volume} {88}},\ \bibinfo {pages} {235103} (\bibinfo {year}
  {2013})}\BibitemShut {NoStop}%
\bibitem [{\citenamefont {You}\ and\ \citenamefont
  {Wen}(2012)}]{youProjectiveNonAbelianStatistics2012a}%
  \BibitemOpen
  \bibfield  {author} {\bibinfo {author} {\bibfnamefont {Y.-Z.}\ \bibnamefont
  {You}}\ and\ \bibinfo {author} {\bibfnamefont {X.-G.}\ \bibnamefont {Wen}},\
  }\bibfield  {title} {\bibinfo {title} {Projective non-{{Abelian Statistics}}
  of {{Dislocation Defects}} in a {{Z}}\_{{N Rotor Model}}},\ }\href
  {https://doi.org/10.1103/PhysRevB.86.161107} {\bibfield  {journal} {\bibinfo
  {journal} {Physical Review B}\ }\textbf {\bibinfo {volume} {86}},\ \bibinfo
  {pages} {161107} (\bibinfo {year} {2012})},\ \Eprint
  {https://arxiv.org/abs/1204.0113} {arXiv:1204.0113 [cond-mat,
  physics:quant-ph]} \BibitemShut {NoStop}%
\bibitem [{\citenamefont {Benhemou}\ \emph {et~al.}(2021)\citenamefont
  {Benhemou}, \citenamefont {Pachos},\ and\ \citenamefont
  {Browne}}]{benhemouNonAbelianStatisticsMixedboundary2021}%
  \BibitemOpen
  \bibfield  {author} {\bibinfo {author} {\bibfnamefont {A.}~\bibnamefont
  {Benhemou}}, \bibinfo {author} {\bibfnamefont {J.~K.}\ \bibnamefont
  {Pachos}},\ and\ \bibinfo {author} {\bibfnamefont {D.~E.}\ \bibnamefont
  {Browne}},\ }\bibfield  {title} {\bibinfo {title} {Non-{{Abelian}} statistics
  with mixed-boundary punctures on the toric code},\ }\href@noop {} {\bibfield
  {journal} {\bibinfo  {journal} {arXiv:2103.08381 [quant-ph]}\ } (\bibinfo
  {year} {2021})},\ \Eprint {https://arxiv.org/abs/2103.08381}
  {arXiv:2103.08381 [quant-ph]} \BibitemShut {NoStop}%
\bibitem [{\citenamefont {Zheng}\ \emph {et~al.}(2015)\citenamefont {Zheng},
  \citenamefont {Dua},\ and\ \citenamefont
  {Jiang}}]{zhengDemonstratingNonAbelianStatistics2015}%
  \BibitemOpen
  \bibfield  {author} {\bibinfo {author} {\bibfnamefont {H.}~\bibnamefont
  {Zheng}}, \bibinfo {author} {\bibfnamefont {A.}~\bibnamefont {Dua}},\ and\
  \bibinfo {author} {\bibfnamefont {L.}~\bibnamefont {Jiang}},\ }\bibfield
  {title} {\bibinfo {title} {Demonstrating non-{{Abelian}} statistics of
  {{Majorana}} fermions using twist defects},\ }\href
  {https://doi.org/10.1103/PhysRevB.92.245139} {\bibfield  {journal} {\bibinfo
  {journal} {Physical Review B}\ }\textbf {\bibinfo {volume} {92}},\ \bibinfo
  {pages} {245139} (\bibinfo {year} {2015})},\ \Eprint
  {https://arxiv.org/abs/1508.04166} {arXiv:1508.04166} \BibitemShut {NoStop}%
\bibitem [{\citenamefont {Brown}\ \emph {et~al.}(2017)\citenamefont {Brown},
  \citenamefont {Laubscher}, \citenamefont {Kesselring},\ and\ \citenamefont
  {Wootton}}]{brownPokingHolesCutting2017a}%
  \BibitemOpen
  \bibfield  {author} {\bibinfo {author} {\bibfnamefont {B.~J.}\ \bibnamefont
  {Brown}}, \bibinfo {author} {\bibfnamefont {K.}~\bibnamefont {Laubscher}},
  \bibinfo {author} {\bibfnamefont {M.~S.}\ \bibnamefont {Kesselring}},\ and\
  \bibinfo {author} {\bibfnamefont {J.~R.}\ \bibnamefont {Wootton}},\
  }\bibfield  {title} {\bibinfo {title} {Poking holes and cutting corners to
  achieve {{Clifford}} gates with the surface code},\ }\href
  {https://doi.org/10.1103/PhysRevX.7.021029} {\bibfield  {journal} {\bibinfo
  {journal} {Physical Review X}\ }\textbf {\bibinfo {volume} {7}},\ \bibinfo
  {pages} {021029} (\bibinfo {year} {2017})},\ \Eprint
  {https://arxiv.org/abs/1609.04673} {arXiv:1609.04673 [cond-mat,
  physics:quant-ph]} \BibitemShut {NoStop}%
\bibitem [{\citenamefont {Acharya}\ \emph {et~al.}(2022)\citenamefont
  {Acharya}, \citenamefont {Aleiner}, \citenamefont {Allen}, \citenamefont
  {Andersen}, \citenamefont {Ansmann}, \citenamefont {Arute}, \citenamefont
  {Arya}, \citenamefont {Asfaw}, \citenamefont {Atalaya}, \citenamefont
  {Babbush}, \citenamefont {Bacon}, \citenamefont {Bardin}, \citenamefont
  {Basso}, \citenamefont {Bengtsson}, \citenamefont {Boixo}, \citenamefont
  {Bortoli}, \citenamefont {Bourassa}, \citenamefont {Bovaird}, \citenamefont
  {Brill}, \citenamefont {Broughton}, \citenamefont {Buckley}, \citenamefont
  {Buell}, \citenamefont {Burger}, \citenamefont {Burkett}, \citenamefont
  {Bushnell}, \citenamefont {Chen}, \citenamefont {Chen}, \citenamefont
  {Chiaro}, \citenamefont {Cogan}, \citenamefont {Collins}, \citenamefont
  {Conner}, \citenamefont {Courtney}, \citenamefont {Crook}, \citenamefont
  {Curtin}, \citenamefont {Debroy}, \citenamefont {Barba}, \citenamefont
  {Demura}, \citenamefont {Dunsworth}, \citenamefont {Eppens}, \citenamefont
  {Erickson}, \citenamefont {Faoro}, \citenamefont {Farhi}, \citenamefont
  {Fatemi}, \citenamefont {Burgos}, \citenamefont {Forati}, \citenamefont
  {Fowler}, \citenamefont {Foxen}, \citenamefont {Giang}, \citenamefont
  {Gidney}, \citenamefont {Gilboa}, \citenamefont {Giustina}, \citenamefont
  {Dau}, \citenamefont {Gross}, \citenamefont {Habegger}, \citenamefont
  {Hamilton}, \citenamefont {Harrigan}, \citenamefont {Harrington},
  \citenamefont {Higgott}, \citenamefont {Hilton}, \citenamefont {Hoffmann},
  \citenamefont {Hong}, \citenamefont {Huang}, \citenamefont {Huff},
  \citenamefont {Huggins}, \citenamefont {Ioffe}, \citenamefont {Isakov},
  \citenamefont {Iveland}, \citenamefont {Jeffrey}, \citenamefont {Jiang},
  \citenamefont {Jones}, \citenamefont {Juhas}, \citenamefont {Kafri},
  \citenamefont {Kechedzhi}, \citenamefont {Kelly}, \citenamefont {Khattar},
  \citenamefont {Khezri}, \citenamefont {Kieferov{\'a}}, \citenamefont {Kim},
  \citenamefont {Kitaev}, \citenamefont {Klimov}, \citenamefont {Klots},
  \citenamefont {Korotkov}, \citenamefont {Kostritsa}, \citenamefont
  {Kreikebaum}, \citenamefont {Landhuis}, \citenamefont {Laptev}, \citenamefont
  {Lau}, \citenamefont {Laws}, \citenamefont {Lee}, \citenamefont {Lee},
  \citenamefont {Lester}, \citenamefont {Lill}, \citenamefont {Liu},
  \citenamefont {Locharla}, \citenamefont {Lucero}, \citenamefont {Malone},
  \citenamefont {Marshall}, \citenamefont {Martin}, \citenamefont {McClean},
  \citenamefont {Mccourt}, \citenamefont {McEwen}, \citenamefont {Megrant},
  \citenamefont {Costa}, \citenamefont {Mi}, \citenamefont {Miao},
  \citenamefont {Mohseni}, \citenamefont {Montazeri}, \citenamefont {Morvan},
  \citenamefont {Mount}, \citenamefont {Mruczkiewicz}, \citenamefont {Naaman},
  \citenamefont {Neeley}, \citenamefont {Neill}, \citenamefont {Nersisyan},
  \citenamefont {Neven}, \citenamefont {Newman}, \citenamefont {Ng},
  \citenamefont {Nguyen}, \citenamefont {Nguyen}, \citenamefont {Niu},
  \citenamefont {O'Brien}, \citenamefont {Opremcak}, \citenamefont {Platt},
  \citenamefont {Petukhov}, \citenamefont {Potter}, \citenamefont {Pryadko},
  \citenamefont {Quintana}, \citenamefont {Roushan}, \citenamefont {Rubin},
  \citenamefont {Saei}, \citenamefont {Sank}, \citenamefont {Sankaragomathi},
  \citenamefont {Satzinger}, \citenamefont {Schurkus}, \citenamefont
  {Schuster}, \citenamefont {Shearn}, \citenamefont {Shorter}, \citenamefont
  {Shvarts}, \citenamefont {Skruzny}, \citenamefont {Smelyanskiy},
  \citenamefont {Smith}, \citenamefont {Sterling}, \citenamefont {Strain},
  \citenamefont {Szalay}, \citenamefont {Torres}, \citenamefont {Vidal},
  \citenamefont {Villalonga}, \citenamefont {Heidweiller}, \citenamefont
  {White}, \citenamefont {Xing}, \citenamefont {Yao}, \citenamefont {Yeh},
  \citenamefont {Yoo}, \citenamefont {Young}, \citenamefont {Zalcman},
  \citenamefont {Zhang},\ and\ \citenamefont
  {Zhu}}]{acharyaSuppressingQuantumErrors2022}%
  \BibitemOpen
  \bibfield  {author} {\bibinfo {author} {\bibfnamefont {R.}~\bibnamefont
  {Acharya}}, \bibinfo {author} {\bibfnamefont {I.}~\bibnamefont {Aleiner}},
  \bibinfo {author} {\bibfnamefont {R.}~\bibnamefont {Allen}}, \bibinfo
  {author} {\bibfnamefont {T.~I.}\ \bibnamefont {Andersen}}, \bibinfo {author}
  {\bibfnamefont {M.}~\bibnamefont {Ansmann}}, \bibinfo {author} {\bibfnamefont
  {F.}~\bibnamefont {Arute}}, \bibinfo {author} {\bibfnamefont
  {K.}~\bibnamefont {Arya}}, \bibinfo {author} {\bibfnamefont {A.}~\bibnamefont
  {Asfaw}}, \bibinfo {author} {\bibfnamefont {J.}~\bibnamefont {Atalaya}},
  \bibinfo {author} {\bibfnamefont {R.}~\bibnamefont {Babbush}}, \bibinfo
  {author} {\bibfnamefont {D.}~\bibnamefont {Bacon}}, \bibinfo {author}
  {\bibfnamefont {J.~C.}\ \bibnamefont {Bardin}}, \bibinfo {author}
  {\bibfnamefont {J.}~\bibnamefont {Basso}}, \bibinfo {author} {\bibfnamefont
  {A.}~\bibnamefont {Bengtsson}}, \bibinfo {author} {\bibfnamefont
  {S.}~\bibnamefont {Boixo}}, \bibinfo {author} {\bibfnamefont
  {G.}~\bibnamefont {Bortoli}}, \bibinfo {author} {\bibfnamefont
  {A.}~\bibnamefont {Bourassa}}, \bibinfo {author} {\bibfnamefont
  {J.}~\bibnamefont {Bovaird}}, \bibinfo {author} {\bibfnamefont
  {L.}~\bibnamefont {Brill}}, \bibinfo {author} {\bibfnamefont
  {M.}~\bibnamefont {Broughton}}, \bibinfo {author} {\bibfnamefont {B.~B.}\
  \bibnamefont {Buckley}}, \bibinfo {author} {\bibfnamefont {D.~A.}\
  \bibnamefont {Buell}}, \bibinfo {author} {\bibfnamefont {T.}~\bibnamefont
  {Burger}}, \bibinfo {author} {\bibfnamefont {B.}~\bibnamefont {Burkett}},
  \bibinfo {author} {\bibfnamefont {N.}~\bibnamefont {Bushnell}}, \bibinfo
  {author} {\bibfnamefont {Y.}~\bibnamefont {Chen}}, \bibinfo {author}
  {\bibfnamefont {Z.}~\bibnamefont {Chen}}, \bibinfo {author} {\bibfnamefont
  {B.}~\bibnamefont {Chiaro}}, \bibinfo {author} {\bibfnamefont
  {J.}~\bibnamefont {Cogan}}, \bibinfo {author} {\bibfnamefont
  {R.}~\bibnamefont {Collins}}, \bibinfo {author} {\bibfnamefont
  {P.}~\bibnamefont {Conner}}, \bibinfo {author} {\bibfnamefont
  {W.}~\bibnamefont {Courtney}}, \bibinfo {author} {\bibfnamefont {A.~L.}\
  \bibnamefont {Crook}}, \bibinfo {author} {\bibfnamefont {B.}~\bibnamefont
  {Curtin}}, \bibinfo {author} {\bibfnamefont {D.~M.}\ \bibnamefont {Debroy}},
  \bibinfo {author} {\bibfnamefont {A.~D.~T.}\ \bibnamefont {Barba}}, \bibinfo
  {author} {\bibfnamefont {S.}~\bibnamefont {Demura}}, \bibinfo {author}
  {\bibfnamefont {A.}~\bibnamefont {Dunsworth}}, \bibinfo {author}
  {\bibfnamefont {D.}~\bibnamefont {Eppens}}, \bibinfo {author} {\bibfnamefont
  {C.}~\bibnamefont {Erickson}}, \bibinfo {author} {\bibfnamefont
  {L.}~\bibnamefont {Faoro}}, \bibinfo {author} {\bibfnamefont
  {E.}~\bibnamefont {Farhi}}, \bibinfo {author} {\bibfnamefont
  {R.}~\bibnamefont {Fatemi}}, \bibinfo {author} {\bibfnamefont {L.~F.}\
  \bibnamefont {Burgos}}, \bibinfo {author} {\bibfnamefont {E.}~\bibnamefont
  {Forati}}, \bibinfo {author} {\bibfnamefont {A.~G.}\ \bibnamefont {Fowler}},
  \bibinfo {author} {\bibfnamefont {B.}~\bibnamefont {Foxen}}, \bibinfo
  {author} {\bibfnamefont {W.}~\bibnamefont {Giang}}, \bibinfo {author}
  {\bibfnamefont {C.}~\bibnamefont {Gidney}}, \bibinfo {author} {\bibfnamefont
  {D.}~\bibnamefont {Gilboa}}, \bibinfo {author} {\bibfnamefont
  {M.}~\bibnamefont {Giustina}}, \bibinfo {author} {\bibfnamefont {A.~G.}\
  \bibnamefont {Dau}}, \bibinfo {author} {\bibfnamefont {J.~A.}\ \bibnamefont
  {Gross}}, \bibinfo {author} {\bibfnamefont {S.}~\bibnamefont {Habegger}},
  \bibinfo {author} {\bibfnamefont {M.~C.}\ \bibnamefont {Hamilton}}, \bibinfo
  {author} {\bibfnamefont {M.~P.}\ \bibnamefont {Harrigan}}, \bibinfo {author}
  {\bibfnamefont {S.~D.}\ \bibnamefont {Harrington}}, \bibinfo {author}
  {\bibfnamefont {O.}~\bibnamefont {Higgott}}, \bibinfo {author} {\bibfnamefont
  {J.}~\bibnamefont {Hilton}}, \bibinfo {author} {\bibfnamefont
  {M.}~\bibnamefont {Hoffmann}}, \bibinfo {author} {\bibfnamefont
  {S.}~\bibnamefont {Hong}}, \bibinfo {author} {\bibfnamefont {T.}~\bibnamefont
  {Huang}}, \bibinfo {author} {\bibfnamefont {A.}~\bibnamefont {Huff}},
  \bibinfo {author} {\bibfnamefont {W.~J.}\ \bibnamefont {Huggins}}, \bibinfo
  {author} {\bibfnamefont {L.~B.}\ \bibnamefont {Ioffe}}, \bibinfo {author}
  {\bibfnamefont {S.~V.}\ \bibnamefont {Isakov}}, \bibinfo {author}
  {\bibfnamefont {J.}~\bibnamefont {Iveland}}, \bibinfo {author} {\bibfnamefont
  {E.}~\bibnamefont {Jeffrey}}, \bibinfo {author} {\bibfnamefont
  {Z.}~\bibnamefont {Jiang}}, \bibinfo {author} {\bibfnamefont
  {C.}~\bibnamefont {Jones}}, \bibinfo {author} {\bibfnamefont
  {P.}~\bibnamefont {Juhas}}, \bibinfo {author} {\bibfnamefont
  {D.}~\bibnamefont {Kafri}}, \bibinfo {author} {\bibfnamefont
  {K.}~\bibnamefont {Kechedzhi}}, \bibinfo {author} {\bibfnamefont
  {J.}~\bibnamefont {Kelly}}, \bibinfo {author} {\bibfnamefont
  {T.}~\bibnamefont {Khattar}}, \bibinfo {author} {\bibfnamefont
  {M.}~\bibnamefont {Khezri}}, \bibinfo {author} {\bibfnamefont
  {M.}~\bibnamefont {Kieferov{\'a}}}, \bibinfo {author} {\bibfnamefont
  {S.}~\bibnamefont {Kim}}, \bibinfo {author} {\bibfnamefont {A.}~\bibnamefont
  {Kitaev}}, \bibinfo {author} {\bibfnamefont {P.~V.}\ \bibnamefont {Klimov}},
  \bibinfo {author} {\bibfnamefont {A.~R.}\ \bibnamefont {Klots}}, \bibinfo
  {author} {\bibfnamefont {A.~N.}\ \bibnamefont {Korotkov}}, \bibinfo {author}
  {\bibfnamefont {F.}~\bibnamefont {Kostritsa}}, \bibinfo {author}
  {\bibfnamefont {J.~M.}\ \bibnamefont {Kreikebaum}}, \bibinfo {author}
  {\bibfnamefont {D.}~\bibnamefont {Landhuis}}, \bibinfo {author}
  {\bibfnamefont {P.}~\bibnamefont {Laptev}}, \bibinfo {author} {\bibfnamefont
  {K.-M.}\ \bibnamefont {Lau}}, \bibinfo {author} {\bibfnamefont
  {L.}~\bibnamefont {Laws}}, \bibinfo {author} {\bibfnamefont {J.}~\bibnamefont
  {Lee}}, \bibinfo {author} {\bibfnamefont {K.}~\bibnamefont {Lee}}, \bibinfo
  {author} {\bibfnamefont {B.~J.}\ \bibnamefont {Lester}}, \bibinfo {author}
  {\bibfnamefont {A.}~\bibnamefont {Lill}}, \bibinfo {author} {\bibfnamefont
  {W.}~\bibnamefont {Liu}}, \bibinfo {author} {\bibfnamefont {A.}~\bibnamefont
  {Locharla}}, \bibinfo {author} {\bibfnamefont {E.}~\bibnamefont {Lucero}},
  \bibinfo {author} {\bibfnamefont {F.~D.}\ \bibnamefont {Malone}}, \bibinfo
  {author} {\bibfnamefont {J.}~\bibnamefont {Marshall}}, \bibinfo {author}
  {\bibfnamefont {O.}~\bibnamefont {Martin}}, \bibinfo {author} {\bibfnamefont
  {J.~R.}\ \bibnamefont {McClean}}, \bibinfo {author} {\bibfnamefont
  {T.}~\bibnamefont {Mccourt}}, \bibinfo {author} {\bibfnamefont
  {M.}~\bibnamefont {McEwen}}, \bibinfo {author} {\bibfnamefont
  {A.}~\bibnamefont {Megrant}}, \bibinfo {author} {\bibfnamefont {B.~M.}\
  \bibnamefont {Costa}}, \bibinfo {author} {\bibfnamefont {X.}~\bibnamefont
  {Mi}}, \bibinfo {author} {\bibfnamefont {K.~C.}\ \bibnamefont {Miao}},
  \bibinfo {author} {\bibfnamefont {M.}~\bibnamefont {Mohseni}}, \bibinfo
  {author} {\bibfnamefont {S.}~\bibnamefont {Montazeri}}, \bibinfo {author}
  {\bibfnamefont {A.}~\bibnamefont {Morvan}}, \bibinfo {author} {\bibfnamefont
  {E.}~\bibnamefont {Mount}}, \bibinfo {author} {\bibfnamefont
  {W.}~\bibnamefont {Mruczkiewicz}}, \bibinfo {author} {\bibfnamefont
  {O.}~\bibnamefont {Naaman}}, \bibinfo {author} {\bibfnamefont
  {M.}~\bibnamefont {Neeley}}, \bibinfo {author} {\bibfnamefont
  {C.}~\bibnamefont {Neill}}, \bibinfo {author} {\bibfnamefont
  {A.}~\bibnamefont {Nersisyan}}, \bibinfo {author} {\bibfnamefont
  {H.}~\bibnamefont {Neven}}, \bibinfo {author} {\bibfnamefont
  {M.}~\bibnamefont {Newman}}, \bibinfo {author} {\bibfnamefont {J.~H.}\
  \bibnamefont {Ng}}, \bibinfo {author} {\bibfnamefont {A.}~\bibnamefont
  {Nguyen}}, \bibinfo {author} {\bibfnamefont {M.}~\bibnamefont {Nguyen}},
  \bibinfo {author} {\bibfnamefont {M.~Y.}\ \bibnamefont {Niu}}, \bibinfo
  {author} {\bibfnamefont {T.~E.}\ \bibnamefont {O'Brien}}, \bibinfo {author}
  {\bibfnamefont {A.}~\bibnamefont {Opremcak}}, \bibinfo {author}
  {\bibfnamefont {J.}~\bibnamefont {Platt}}, \bibinfo {author} {\bibfnamefont
  {A.}~\bibnamefont {Petukhov}}, \bibinfo {author} {\bibfnamefont
  {R.}~\bibnamefont {Potter}}, \bibinfo {author} {\bibfnamefont {L.~P.}\
  \bibnamefont {Pryadko}}, \bibinfo {author} {\bibfnamefont {C.}~\bibnamefont
  {Quintana}}, \bibinfo {author} {\bibfnamefont {P.}~\bibnamefont {Roushan}},
  \bibinfo {author} {\bibfnamefont {N.~C.}\ \bibnamefont {Rubin}}, \bibinfo
  {author} {\bibfnamefont {N.}~\bibnamefont {Saei}}, \bibinfo {author}
  {\bibfnamefont {D.}~\bibnamefont {Sank}}, \bibinfo {author} {\bibfnamefont
  {K.}~\bibnamefont {Sankaragomathi}}, \bibinfo {author} {\bibfnamefont
  {K.~J.}\ \bibnamefont {Satzinger}}, \bibinfo {author} {\bibfnamefont {H.~F.}\
  \bibnamefont {Schurkus}}, \bibinfo {author} {\bibfnamefont {C.}~\bibnamefont
  {Schuster}}, \bibinfo {author} {\bibfnamefont {M.~J.}\ \bibnamefont
  {Shearn}}, \bibinfo {author} {\bibfnamefont {A.}~\bibnamefont {Shorter}},
  \bibinfo {author} {\bibfnamefont {V.}~\bibnamefont {Shvarts}}, \bibinfo
  {author} {\bibfnamefont {J.}~\bibnamefont {Skruzny}}, \bibinfo {author}
  {\bibfnamefont {V.}~\bibnamefont {Smelyanskiy}}, \bibinfo {author}
  {\bibfnamefont {W.~C.}\ \bibnamefont {Smith}}, \bibinfo {author}
  {\bibfnamefont {G.}~\bibnamefont {Sterling}}, \bibinfo {author}
  {\bibfnamefont {D.}~\bibnamefont {Strain}}, \bibinfo {author} {\bibfnamefont
  {M.}~\bibnamefont {Szalay}}, \bibinfo {author} {\bibfnamefont
  {A.}~\bibnamefont {Torres}}, \bibinfo {author} {\bibfnamefont
  {G.}~\bibnamefont {Vidal}}, \bibinfo {author} {\bibfnamefont
  {B.}~\bibnamefont {Villalonga}}, \bibinfo {author} {\bibfnamefont {C.~V.}\
  \bibnamefont {Heidweiller}}, \bibinfo {author} {\bibfnamefont
  {T.}~\bibnamefont {White}}, \bibinfo {author} {\bibfnamefont
  {C.}~\bibnamefont {Xing}}, \bibinfo {author} {\bibfnamefont {Z.~J.}\
  \bibnamefont {Yao}}, \bibinfo {author} {\bibfnamefont {P.}~\bibnamefont
  {Yeh}}, \bibinfo {author} {\bibfnamefont {J.}~\bibnamefont {Yoo}}, \bibinfo
  {author} {\bibfnamefont {G.}~\bibnamefont {Young}}, \bibinfo {author}
  {\bibfnamefont {A.}~\bibnamefont {Zalcman}}, \bibinfo {author} {\bibfnamefont
  {Y.}~\bibnamefont {Zhang}},\ and\ \bibinfo {author} {\bibfnamefont
  {N.}~\bibnamefont {Zhu}},\ }\href {https://doi.org/10.48550/arXiv.2207.06431}
  {\bibinfo {title} {Suppressing quantum errors by scaling a surface code
  logical qubit}} (\bibinfo {year} {2022}),\ \Eprint
  {https://arxiv.org/abs/2207.06431} {arXiv:2207.06431 [quant-ph]} \BibitemShut
  {NoStop}%
\bibitem [{Note1()}]{Note1}%
  \BibitemOpen
  \bibinfo {note} {As a consequence of the ``handshaking lemma'' that every
  graph has an even number of odd degree vertices\cite
  {eulerSolutioProblematisAd1741}}\BibitemShut {NoStop}%
\bibitem [{Note2()}]{Note2}%
  \BibitemOpen
  \bibinfo {note} {On a general manifold there may be relations amongst the
  stabilizers that depend on topology and boundary conditions which can
  increase the effective code subspace.}\BibitemShut {Stop}%
\bibitem [{Note3()}]{Note3}%
  \BibitemOpen
  \bibinfo {note} {In other words, an anyon with the quantum dimension
  $\protect \sqrt {2}$.}\BibitemShut {Stop}%
\bibitem [{\citenamefont
  {Kasteleyn}(1963)}]{kasteleynDimerStatisticsPhase1963}%
  \BibitemOpen
  \bibfield  {author} {\bibinfo {author} {\bibfnamefont {P.~W.}\ \bibnamefont
  {Kasteleyn}},\ }\bibfield  {title} {\bibinfo {title} {Dimer {{Statistics}}
  and {{Phase Transitions}}},\ }\href {https://doi.org/10.1063/1.1703953}
  {\bibfield  {journal} {\bibinfo  {journal} {Journal of Mathematical Physics}\
  }\textbf {\bibinfo {volume} {4}},\ \bibinfo {pages} {287} (\bibinfo {year}
  {1963})}\BibitemShut {NoStop}%
\bibitem [{Note4()}]{Note4}%
  \BibitemOpen
  \bibinfo {note} {Kasteleyn structures were introduced to study dimer
  models\cite {kasteleynDimerStatisticsPhase1963}, and were later related to 2D
  spin structures\cite
  {cimasoniDimersSurfaceGraphs2007,cimasoniDimersSurfaceGraphs2008}}\BibitemShut
  {NoStop}%
\bibitem [{Note5()}]{Note5}%
  \BibitemOpen
  \bibinfo {note} {\label {fn:globalphase}More precisely, up to a global phase,
  which for us is irrelevant.}\BibitemShut {Stop}%
\bibitem [{Note6()}]{Note6}%
  \BibitemOpen
  \bibinfo {note} {Note that we have given the analogous condition, an odd
  number of clockwise arrows in each plaquette, for the Kasteleyn
  orientation.}\BibitemShut {Stop}%
\bibitem [{Note7()}]{Note7}%
  \BibitemOpen
  \bibinfo {note} {In general, the same would be true if we fixed the parity of
  clockwise edges across all stabilizer plaquettes.}\BibitemShut {Stop}%
\bibitem [{\citenamefont {Fradkin}(2013)}]{fradkinFieldTheoriesCondensed2013}%
  \BibitemOpen
  \bibfield  {author} {\bibinfo {author} {\bibfnamefont {E.}~\bibnamefont
  {Fradkin}},\ }\href {https://doi.org/10.1017/CBO9781139015509} {\emph
  {\bibinfo {title} {Field {{Theories}} of {{Condensed Matter Physics}}}}},\
  \bibinfo {edition} {2nd}\ ed.\ (\bibinfo  {publisher} {{Cambridge University
  Press}},\ \bibinfo {address} {{Cambridge}},\ \bibinfo {year}
  {2013})\BibitemShut {NoStop}%
\bibitem [{Note8()}]{Note8}%
  \BibitemOpen
  \bibinfo {note} {More precisely, we will find violations of different type
  are created in pairs at the ends of string operators.}\BibitemShut {Stop}%
\bibitem [{\citenamefont {Wilson}(1974)}]{wilsonConfinementQuarks1974}%
  \BibitemOpen
  \bibfield  {author} {\bibinfo {author} {\bibfnamefont {K.~G.}\ \bibnamefont
  {Wilson}},\ }\bibfield  {title} {\bibinfo {title} {Confinement of quarks},\
  }\href {https://doi.org/10.1103/PhysRevD.10.2445} {\bibfield  {journal}
  {\bibinfo  {journal} {Physical Review D}\ }\textbf {\bibinfo {volume} {10}},\
  \bibinfo {pages} {2445} (\bibinfo {year} {1974})}\BibitemShut {NoStop}%
\bibitem [{\citenamefont {{'t
  Hooft}}(1978)}]{thooftPhaseTransitionPermanent1978}%
  \BibitemOpen
  \bibfield  {author} {\bibinfo {author} {\bibfnamefont {G.}~\bibnamefont {{'t
  Hooft}}},\ }\bibfield  {title} {\bibinfo {title} {On the phase transition
  towards permanent quark confinement},\ }\href
  {https://doi.org/10.1016/0550-3213(78)90153-0} {\bibfield  {journal}
  {\bibinfo  {journal} {Nuclear Physics B}\ }\textbf {\bibinfo {volume}
  {138}},\ \bibinfo {pages} {1} (\bibinfo {year} {1978})}\BibitemShut {NoStop}%
\bibitem [{Note9()}]{Note9}%
  \BibitemOpen
  \bibinfo {note} {Equivalently, since 't Hooft loops measure \(\varepsilon \)
  parity and deform to canonical augmented Wilson lines.}\BibitemShut {Stop}%
\bibitem [{Note10()}]{Note10}%
  \BibitemOpen
  \bibinfo {note} {More complicated braids can achieve larger code distances by
  constant factors in certain cases.}\BibitemShut {Stop}%
\bibitem [{Note11()}]{Note11}%
  \BibitemOpen
  \bibinfo {note} {The precise procedure to accomplish this depends slightly on
  the available geometry when the devices are small, for example in many cases
  it is convenient to modify the step from \(t=2\) to \(t=3\) by moving edges
  to the \protect \emph {right} instead of the left of the anyon to move it
  upwards. Such an extension is shown in Appendix A.}\BibitemShut {Stop}%
\bibitem [{\citenamefont {Greenberger}\ \emph {et~al.}(2007)\citenamefont
  {Greenberger}, \citenamefont {Horne},\ and\ \citenamefont
  {Zeilinger}}]{greenbergerGoingBellTheorem2007}%
  \BibitemOpen
  \bibfield  {author} {\bibinfo {author} {\bibfnamefont {D.~M.}\ \bibnamefont
  {Greenberger}}, \bibinfo {author} {\bibfnamefont {M.~A.}\ \bibnamefont
  {Horne}},\ and\ \bibinfo {author} {\bibfnamefont {A.}~\bibnamefont
  {Zeilinger}},\ }\href {https://doi.org/10.48550/arXiv.0712.0921} {\bibinfo
  {title} {Going {{Beyond Bell}}'s {{Theorem}}}} (\bibinfo {year} {2007}),\
  \Eprint {https://arxiv.org/abs/0712.0921} {arXiv:0712.0921 [quant-ph]}
  \BibitemShut {NoStop}%
\bibitem [{\citenamefont {Bravyi}\ \emph {et~al.}(2018)\citenamefont {Bravyi},
  \citenamefont {Englbrecht}, \citenamefont {Koenig},\ and\ \citenamefont
  {Peard}}]{bravyiCorrectingCoherentErrors2018}%
  \BibitemOpen
  \bibfield  {author} {\bibinfo {author} {\bibfnamefont {S.}~\bibnamefont
  {Bravyi}}, \bibinfo {author} {\bibfnamefont {M.}~\bibnamefont {Englbrecht}},
  \bibinfo {author} {\bibfnamefont {R.}~\bibnamefont {Koenig}},\ and\ \bibinfo
  {author} {\bibfnamefont {N.}~\bibnamefont {Peard}},\ }\bibfield  {title}
  {\bibinfo {title} {Correcting coherent errors with surface codes},\ }\href
  {https://doi.org/10.1038/s41534-018-0106-y} {\bibfield  {journal} {\bibinfo
  {journal} {npj Quantum Information}\ }\textbf {\bibinfo {volume} {4}},\
  \bibinfo {pages} {55} (\bibinfo {year} {2018})},\ \Eprint
  {https://arxiv.org/abs/1710.02270} {arXiv:1710.02270 [quant-ph]} \BibitemShut
  {NoStop}%
\bibitem [{Note12()}]{Note12}%
  \BibitemOpen
  \bibinfo {note} {In our system of Ising anyons Clifford gates can be
  implemented fault-tolerantly. A non-Clifford $T$-gate necessary for universal
  computation can be constructed by replacing $\pi /4\to \pi /8$ in Eq.~(\ref
  {eq:gauge-invariant-swap}) and taking the line between any two anyons. This
  operation is not fault-tolerant.}\BibitemShut {Stop}%
\bibitem [{\citenamefont {Cimasoni}\ and\ \citenamefont
  {Reshetikhin}(2007)}]{cimasoniDimersSurfaceGraphs2007}%
  \BibitemOpen
  \bibfield  {author} {\bibinfo {author} {\bibfnamefont {D.}~\bibnamefont
  {Cimasoni}}\ and\ \bibinfo {author} {\bibfnamefont {N.}~\bibnamefont
  {Reshetikhin}},\ }\bibfield  {title} {\bibinfo {title} {Dimers on surface
  graphs and spin structures. {{I}}},\ }\href
  {https://doi.org/10.1007/s00220-007-0302-7} {\bibfield  {journal} {\bibinfo
  {journal} {Communications in Mathematical Physics}\ }\textbf {\bibinfo
  {volume} {275}},\ \bibinfo {pages} {187} (\bibinfo {year} {2007})},\ \Eprint
  {https://arxiv.org/abs/math-ph/0608070} {arXiv:math-ph/0608070} \BibitemShut
  {NoStop}%
\bibitem [{\citenamefont {Cimasoni}\ and\ \citenamefont
  {Reshetikhin}(2008)}]{cimasoniDimersSurfaceGraphs2008}%
  \BibitemOpen
  \bibfield  {author} {\bibinfo {author} {\bibfnamefont {D.}~\bibnamefont
  {Cimasoni}}\ and\ \bibinfo {author} {\bibfnamefont {N.}~\bibnamefont
  {Reshetikhin}},\ }\bibfield  {title} {\bibinfo {title} {Dimers on surface
  graphs and spin structures. {{II}}},\ }\href
  {https://doi.org/10.1007/s00220-008-0488-3} {\bibfield  {journal} {\bibinfo
  {journal} {Communications in Mathematical Physics}\ }\textbf {\bibinfo
  {volume} {281}},\ \bibinfo {pages} {445} (\bibinfo {year} {2008})},\ \Eprint
  {https://arxiv.org/abs/0704.0273} {arXiv:0704.0273 [math-ph]} \BibitemShut
  {NoStop}%
\bibitem [{\citenamefont {Euler}(1741)}]{eulerSolutioProblematisAd1741}%
  \BibitemOpen
  \bibfield  {author} {\bibinfo {author} {\bibfnamefont {L.}~\bibnamefont
  {Euler}},\ }\bibfield  {title} {\bibinfo {title} {Solutio problematis ad
  geometriam situs pertinentis},\ }\href@noop {} {\bibfield  {journal}
  {\bibinfo  {journal} {Commentarii academiae scientiarum Petropolitanae}\ ,\
  \bibinfo {pages} {128}} (\bibinfo {year} {1741})}\BibitemShut {NoStop}%
\end{thebibliography}%

\appendix

\section{Appendix}

\section{A. Examples on a \(5 \times 5\) qubit system}
\label{sec:examples-5-times}

In this section, we illustrate some of the above steps explicitly in a
\(5 \times 5\) qubit system. First, we specify the initial state by giving
appropriate stabilizers, see Fig.~5(a). The initial PSC is simply a square
graph in the bulk, with 4 anyons at the corners: this is just a surface code
encoding 1 qubit \cite{bravyiQuantumCodesLattice1998}, and has previously
been prepared on a superconducting quantum
processor\cite{satzingerRealizingTopologicallyOrdered2021}. We assume the
state can be prepared so that the vertical 't Hooft line shown, whose
explicit form is also given, takes a definite value.

Next, in Fig~5(b) we show the Pauli string that generates a motion that
appears in the middle of a possible extension of the braid in Fig.~4(g) to a
larger system. According to the rules from the main text, we use
\Cref{eq:gauge-invariant-swap} with \(U_{+}\) to perform this move. As
discussed in the main text, if we wished to move the composite of an anyon
with attached \(\mathbb{Z}_2^{(s)}\) flux, we would use \(U_{-}\).

\begin{figure*}[htbp]
  \centering
  \includegraphics{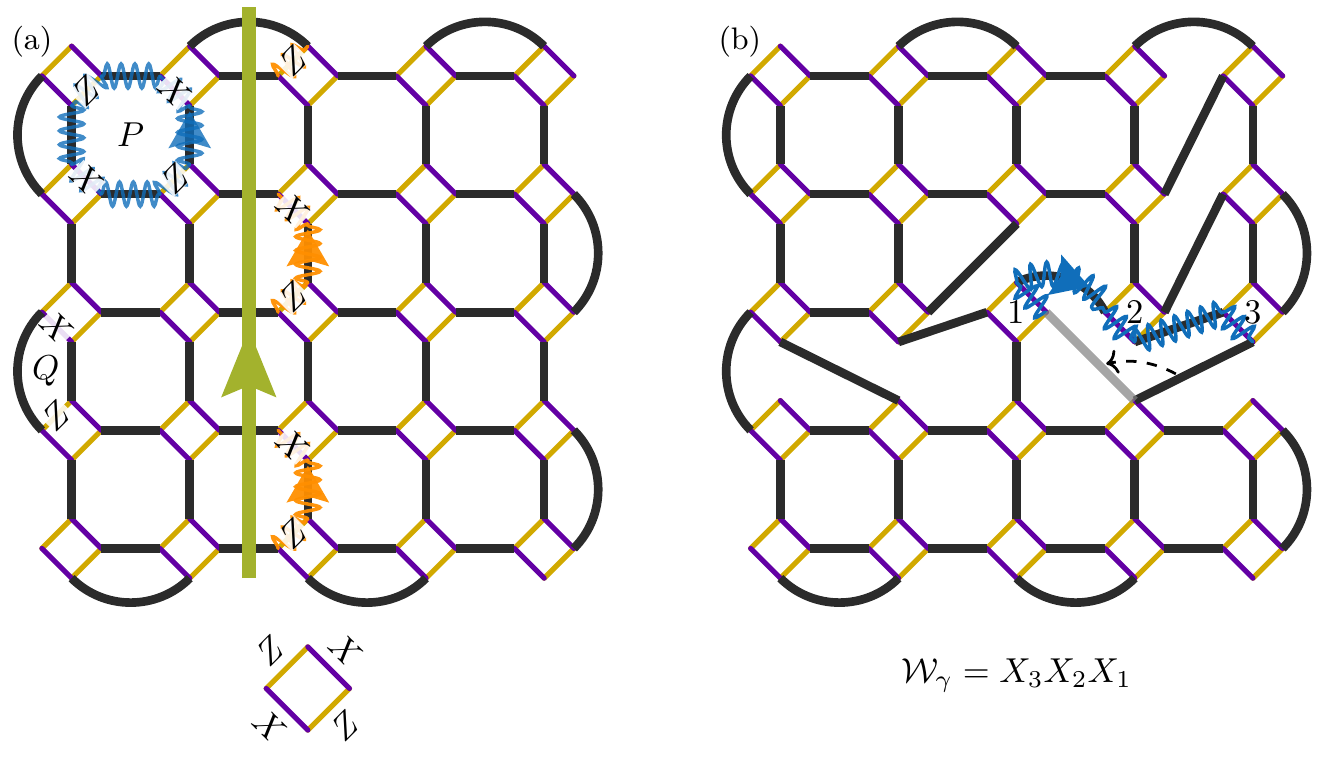}
  \caption[]{%
    \begin{captionlist}
      \item The surface code as a PSC, with several visualizations of \Cref{eq:aW-pauli}. We show the two stabilizers associated to the plaquettes \(P\) and \(Q\), as well as the Wilson line for \(B(P)\). We also apply the rules from Fig. 2(g) to construct the Pauli string for the logical \(Z\) 't Hooft line (the Pauli string appears to the right of the line).
      \item The Pauli string that generates a movement circuit, associated to the Wilson line shown. This is another direct application of \Cref{eq:aW-pauli}.
    \end{captionlist}%
  }
\end{figure*}

\section{B. Kasteleyn orientations and path deformations}
\label{sec:kasteleyn-appendix}
A Kasteleyn orientation \autocite{kasteleynDimerStatisticsPhase1963} always
exists on a surface graph with an even number of vertices
\autocite{cimasoniDimersSurfaceGraphs2007,cimasoniDimersSurfaceGraphs2008}. There
is a precise sense in which such an orientation behaves like a typical
\(\mathbb{Z}_{2}\) gauge field. One Kasteleyn orientation can be taken to
\emph{any} other by flipping arrows on links crossed by cycles through the dual
graph
\autocite{cimasoniDimersSurfaceGraphs2007,cimasoniDimersSurfaceGraphs2008},
with contractible cycles generated by the \(\mathbb{Z}_2^{(K)}\)
transformation described in the main text. This is the same way that a
conventional \(\mathbb{Z}_2\) field configuration can be taken to any other
with the same pattern of local flux (the transformations corresponding to the
contractible loops are gauge). The reason is that any cycle flips an even
number of arrows in each plaquette.

We will re-use the definition \(W^{(K)}_{\gamma}\) from \Cref{eq:WsK} for directed
paths and loops in a general graph (i.e. it is \(+1\) (\(-1\)) if there are
an even (odd) number of arrows along the path that point opposite the
direction of the path). Importantly, the above discussion proves that for a
directed \emph{contractible} loop \(\gamma\), \(W^{(K)}_{\gamma}\) is independent of
the choice of Kasteleyn orientation. The invariant value can be understood by
contracting a simple (no self-intersections) counter-clockwise loop to a
single face \(F_0\), where the definition of a Kasteleyn orientation is that
\(W^{(K)}_{\partial F_0} = -1\). To do this, we give a useful rule for ``pushing''
segments of a path \(\gamma\) through a face \(F\). Part of \(\partial F\) is
\(\gamma_1 \subset \gamma\). The complement of \(\gamma_1\) in the boundary
\(\partial F\) of the face \(F\) is \(\gamma_2\). To ``deform'' the path
\(\gamma\) is to replace \(\gamma_1\) with \(\gamma_2\) (in the same direction), obtaining a
path \(\gamma'\). To compute the accompanying sign change
\(W^{(K)}_{\gamma} (W^{(K)}_{\gamma'})^{-1}\), note that \(\gamma_2\) traverses
\(F\) clockwise, so the Kasteleyn condition for \(F\) refers to the reversed
path \(\hat{R} \gamma_2\). We have the general formula
\(W_{\hat{R}\gamma}^{(K)} = (-1)^{\pathlen(\gamma)} W_{\gamma}^{(K)} = -
(-1)^{\pathverts(\gamma)} W_{\gamma}^{(K)}\). Combining with the Kasteleyn condition,
we find in the case of deformation through a single face
\begin{equation}
  \label{eq:kasteleyn-single-face-deformation}
  W^{(K)}_{\gamma_1} = (-1)^{\pathverts(\gamma_2)-2} W_{\gamma_2}^{(K)}. 
\end{equation}
But \(\pathverts(\gamma_{2})-2\) simply counts the number of vertices that
are in the interior of \(\gamma\) but not \(\gamma'\). Continuing this way
until we arrive at a single face, calling \(V_B(\gamma)\) the number of
vertices in the \emph{interior} of the loop \(\gamma\) we find
\begin{equation*}
  W^{(K)}_{\gamma} = - (-1)^{V_B(\gamma)}
\end{equation*}
for any counter-clockwise simple loop \(\gamma\).
\subsection{B.1 Canonical paths and loops}
\label{sec:canonical-kasteleyn-paths-loops}
We now return to the special case of the decorated PSC graph, and always
focus on a disk-like region. Every simple counter-clockwise loop in the
undecorated graph, \(\hat{\gamma}\), could naturally correspond to
\(2^{\pathlen(\hat{\gamma})}\) directed loops through the decorated graph,
because at each added diamond we can choose whether to go around it clockwise
or counter-clockwise.

For open paths we also choose which diamond vertex the path ends at. The
physical requirement of \(\mathbb{Z}^{(s)}_2\) invariance for the augmented
Wilson lines that are built from this path constrains it to end on a
different vertex of the endpoint diamonds than where it entered. This is the
definition of a \emph{valid} path.

In fact, we can see by inspection of \Cref{eq:aW,eq:WsK} that the choice of
how diamonds are traversed only affects the Kasteleyn part of a loop or line,
and therefore simply changes the sign of the operator. Moreover, by the
deformation formula \Cref{eq:kasteleyn-single-face-deformation} we see that
if a line touches a diamond an odd number of times, it does not matter which
way we traverse that diamond. Thus we only have to keep careful track of
``wedges'' where a diamond is touched precisely twice in a row. When building
various operators this can simply be chosen as convenient (c.f. the movement
gates in Fig. 4(d-f)), but to predict braiding outcomes by deformation of
Wilson lines we need to know which way to take the wedges. Remarkably,
unitarity, locality, and gauge invariance determine that we can always take
Wilson lines with wedges pointing to the \emph{right} (i.e. traversing the
diamond counter-clockwise) to measure fusion outcomes. In the main text, to
give a more concise definition of canonicity we simply insisted on all lines
traversing the diamonds counter-clockwise, which is equivalent to the
definition here. The more refined definition here is convenient for various
proofs since fewer cases need to be checked. Note in particular
\((-1)^{N_{ll}(\gamma)}\) (defined below \Cref{eq:aW-pauli}) only depends on the
number of wedges.

Consider now a simple canonical loop \(\gamma\), and cut away the exterior edges
and vertices, so that \(\gamma\) becomes the boundary of a graph
\(\tilde{G}\). The important geometric property of a canonical loop is that,
when viewed as the boundary of \(\tilde{G}\), \(\gamma\) has an even number of odd
degree vertices. By the ``handshaking lemma''
\autocite{eulerSolutioProblematisAd1741}, this means that the number of
odd-degree vertices on the interior of the loop is even. The only even-degree
vertices are the unpaired ones, so \(V_B(\gamma) = N_{\sigma}(\gamma)\) and for a
contractible canonical loop
\begin{equation*}
  \label{eq:canonical-loop-proved-kasteleyn}
  W^{(K)}_{\gamma} = - (-1)^{N_{\sigma}(\gamma)}.
\end{equation*}
This proves the results \Cref{eq:N_sigma,eq:augmented-wilson-loop-flux}.

To prove \Cref{eq:wilson-line-deformation}, we note that, as usual, the ratio
of the Kasteleyn Wilson lines, \(W^{(K)}_{\gamma} (W^{(K)}_{\gamma'})^{-1}\) of two
valid canonical paths \(\gamma, \gamma'\) with endpoints at the same two anyons 1,2 is
a product of ratios of canonical paths that form simple closed loops. Each
loop consists of two segments, one from \(\gamma\) and one from \(\gamma'\). We only
need to consider one such loop. One of the segments is counter-clockwise
about the loop and the other clockwise; call the counter-clockwise segment
\(\gamma_1\) and the other \(\gamma_2\). In fact, by enumerating the ways in which
canonical paths can split from each other, one finds that \(\gamma_2\) is
\emph{always} valid (this is not necessarily the case for \(\gamma_1\)). The
reversed path \(\hat{R} \gamma_2\) may not be canonical, and
\(W_{\hat{R}(\gamma_2)}^{(K)} = - (-1)^{N_{ll}(\gamma_{2})} W_{\gamma_2}^{(K)}\). To make
the reversed path canonical, each wedge should be flipped, which cancels the
factor \((-1)^{N_{ll}(\gamma_2)}\); call this path \(R \gamma_2\). The path
\(\gamma = \gamma_1 \cup R\gamma_2\) is now a canonical loop, and we find
\begin{equation*}
  W_{\gamma_1}^{(K)} (W_{\gamma_2}^{(K)})^{-1} = - W_{\gamma}^{(K)} = (-1)^{N_{\sigma}(\gamma)} \equiv (-1)^{N_{\sigma}(\gamma_1, \gamma_2)}.
\end{equation*}
This expression gives a precise definition of \(N_{\sigma}(\gamma,\gamma')\) in the main
text, which is only necessary when the Wilson lines pass directly through
unpaired anyons away from the endpoints (the latter are of course in common,
and it is straightforward to check that they never contribute to this flux
difference). In practice, if there are few anyons on the Wilson line it is
often simpler to deform the line by one plaquette using
\Cref{eq:kasteleyn-single-face-deformation} first, and then apply the
counting formula. We also note that because of some exceptions at the
endpoints, in the main text we only stated \Cref{eq:wilson-line-deformation}
for Wilson lines between anyons. The formula also applies with other
conditions, most obviously when the paths \(\gamma, \gamma'\) differ only away from
their endpoint diamonds.

\end{document}